\documentclass[aps,pre,preprint,groupedaddress,nofootinbib,showpacs,onecolumn]{revtex4-1}

\usepackage[english]{babel}
\usepackage{bbm}
\usepackage{amssymb}
\usepackage{amsmath,amsfonts,amsthm,bm} 
\usepackage{mathtools}
\usepackage{times}
\usepackage{bm}
\usepackage{natbib}
\usepackage{url}
\usepackage{xcolor}
\usepackage{multirow}
\usepackage{tabularx, ragged2e, booktabs}
\newcolumntype{Y}{>{\centering\arraybackslash}X}
\usepackage{adjustbox}
\usepackage{cprotect} 

\begin{document}

\title{Stochastic Resonance for Non-Equilibrium Systems}

\author{Valerio Lucarini}
\affiliation{Centre for the Mathematics of Planet Earth, University of Reading, Reading, UK}
\affiliation{Department of Mathematics and Statistics, University of Reading, Reading UK}
\affiliation{CEN, University of Hamburg, Hamburg, Germany}
 
\email[]{v.lucarini@reading.ac.uk}
\date{\today}
\begin{abstract}

Stochastic resonance (SR) is a  prominent phenomenon in many natural and engineered noisy system, whereby the response to a periodic forcing is greatly amplified when the intensity of the noise is tuned to within a specific range of values. We propose here a general mathematical framework based on large deviation theory, and, specifically, on the theory of quasi-potentials, for describing SR in noisy N-dimensional non-equilibrium systems possessing two metastable states and undergoing a periodically modulated forcing. The drift and the volatility fields of the equations of motion can be fairly general and the competing attractors of the deterministic dynamics and the edge state living on the basin boundary can, in principle, feature chaotic dynamics. Similarly, the perturbation field of the forcing can be fairly general. Our approach is able to recover as special cases the classical results previously presented in the literature for systems obeying detailed balance and allows for expressing the parameters describing SR  and the statistics of residence times in the  two-state approximation in terms of the unperturbed drift field, the volatility field, and the perturbation field. We clarify which specific properties of the forcing are relevant for amplifying or suppressing SR in a system, and classify forcings according to classes of equivalence. Our results indicate a route for a detailed understanding of SR in rather general  systems.

\end{abstract}


\maketitle

\section{Introduction}
Stochastic resonance (SR) is a rather special and somewhat counterintuitive mechanism where noise plays the constructive role of catalyzing the amplification of the response of a system to a weak periodic signal. SR was originally proposed independently by  Benzi \textit{et al.} \cite{Benzi1981,Benzi1982,Benzi1983} and by \citet{Nicolis1981,Nicolis1982} 
as a way to explain the occurrence of periodically spaced ice ages in the Quaternary era despite the presence of extremely weak periodic modulations of the incoming solar radiation due to the Milankovich cycles. Since then, SR has been found and studied in a myriad of natural and engineered systems, and has been thoroughly explored through theory, experiments, and numerical simulations. We report examples from laser systems \cite{McNamara1988}, atomic physics \cite{Stroescu2016}, nanostructures \cite{Wagner2019}, optics \cite{Schiavoni2002}, control theory \cite{Jerome2014}, circuits \cite{Korneta2006}, ecology \cite{Han2014}, geosciences \cite{Alley2001,Ganopolski2002}, biology \cite{Hanggi2002},  physiology \cite{Lalwani2019}, neurosciences \cite{,McDonnell2008}, and psychology \cite{Ward2007}, among others. Many valuable reviews on the topic are available \cite{Gammaitoni1998,Anishchenko1999,Wellens2003,Benzi2010}. Additionally, SR has become the subject of careful mathematical investigations; see, e.g., \cite{Freidlin2000,Herrmann2005,Imkeller2002,Imkeller2004,Herrmann2014}.

The mathematical archetype  for  SR is the system whose (overdamped) dynamics is described by the following stochastic differential equation (SDE):
\begin{equation}
d x/ dt = -V'(x) +\epsilon \cos(\omega t)+ \sigma dW/dt \label{eq1}
\end{equation}
where $V(x)=-a x^2+ b x^4$ ($a, b>0$) features two stable equilibria at $x=\pm x_0=\pm(a/2b)^{1/2}$, $\omega$ is the frequency of the periodic forcing, $dW$ is the increment of a brownian motion, and $\sigma$ modulates the intensity of the stochastic forcing. Stochastic forcing will lead to trajectories performing transitions between the basin of attractions of the two stable equilibria; we indicate by $2$ ($1$) the basin of attraction of $x_0$ ($-x_0$). If $\epsilon=0$ and $\sigma>0$, the classical Kramers' theory \cite{Kramers1940}  predicts that, in the weak-noise limit, the average transition rate between the two basins of attraction is 
\begin{eqnarray}
r_{2,\sigma}&=r_{1,\sigma}=\frac{1}{2\pi}|V''(x)|_0||V''(x)|_{x_0}|\exp\left(-2\frac{V(0)-V(x_0)}{\sigma^2}\right)\nonumber\\
&=\frac{4}{\pi}a^2\exp\left(-\frac{a^2/2b}{\sigma^2}\right)\label{kramers}
\end{eqnarray}
The Kramers' formula has been generalised by by \citet{Bovier2004} for N-dimensional gradient flows of the form ${{d{\bm{x}}}(t)}=-\bm{\nabla} U(\bm{x})dt+\sigma d{\bm{W}}$, where $\bm{x} \in \mathbb{R}^{N}$ and $d{\bm{W}}$ is a vector whose components are N independent increments of a brownian motion; see also \cite{Berglund2013}.  

The classical result on SR says that, by and large, if we now switch on the periodic forcing by setting $\epsilon>0$, one gets that the periodic component of the expectation value $\langle x(t) \rangle$ is greatly amplified if $r_{1,\sigma}=r_{2,\sigma}\approx \omega/\pi=2/T$. By tuning the noise in this way, one can obtain a virtually perfect synchronization between the periodic forcing and the transitions between the two basins of attraction for each individual ensemble member; see also \cite{Fox1993}. The problem has been later generalised to the case of more general  noise laws \cite{Tessone1998,Jia2000,Jia2001,Kuhwald2016}, while a  general treatment of SR in an asymmetric potential with complex stochastic forcing has been presented by \citet{Qiao2016}. 

A useful simplification to the problem is obtained by performing a maximal coarse graining procedure such that the system is reduced to two discrete states, which corresponding to the stable equilibria in  the continuum description \cite{McNamara1989,Gammaitoni1998}. The coarse graining leads to concentrating on the interwell hopping, and to neglecting the intrawell dynamics. Solid mathematical foundations to this approach can be found in, e.g, \cite{Lelievre2015,Gesu2019}. Let's refer to these states as $x_1$ and $x_2$, corresponding to the two basin of attractions of $x_0$ and $-x_0$, respectively. 
The analysis of SR for the two-state model has been presented for the symmetric case 
by, e.g, \cite{McNamara1989, Gammaitoni1998}, while general results have been presented for the asymmetric case and for non-Gaussian stochastic forcing in \cite{Bouzat1999,Wio1999}. We will come back to these results later in the paper.

Most of the result on SR have been derived in the case of one-dimensional systems or, more generally, of N-dimensional gradient flows. The goal of this paper is propose a general formulation of SR able to encompass the case of non-equilibrium systems possessing two metastable states.  The classical results valid for system obeying detailed balance are obtained as special cases. We will then consider N-dimensional stochastic differential equations (SDEs) with a fairly general class of noise laws, and assume that, if the noise is switched off, the deterministic dynamics features two competing asymptotic states.  We will rewrite some of the classical results of SR - within the framework of the two-state approximation - in terms of quantities that can be derived from the equations of motion. Our treatment will in principle include the case of stochastically perturbed systems featuring, when noise is removed, two competing  chaotic attractors supported on strange sets. The  edge state embedded in the basin boundary\cite{Grebogi1983,Robert2000,Ott2002,Vollmer2009,LucariniBodai2017} can as well, in principle, be  supported  on a strange set and feature chaotic dynamics. While some hints at SR for this general case have been proposed in the literature \cite{Anishchenko1999}, a complete treatment has not been yet presented, as for the author's knowledge. 

Note that different mechanisms of SR-like phenomena for chaotic systems have been discussed in the literature, where deterministic chaos plays the role of internally generate noise, and no external stochastic forcing is needed. Resonant response to periodic forcing has been found in the case of systems inhabiting preferentially two distinct preferred regions of phase space \cite{Nicolis1993,Anishchenko1993}, or in the case a parameter periodically fluctuates below and above the value determining the onset of chaotic motions \cite{Crisanti1994}.

The paper is structured as follows. In Sect. \ref{mathframe} we present the mathematical framework we deem useful for studying N-dimensional noisy non-equilibrium systems. In Sect. \ref{StoRes} we provide a derivation of SR conditions based upon a linear response approach, and study the resonant amplification of the system's response to a periodic forcing of very general nature. We draw our conclusion, discuss the limitations of the present work, and present possible future lines of research in Sect. \ref{conclu}. Appendix \ref{residence} contains results pertaining the statistics of residence times, i.e. the time intervals  spent consecutively in each state before a noise-induced transition takes place.

\section{Mathematical Framework}\label{mathframe}
We consider an SDE in It\^o form written as
\begin{equation}\label{eqapp}
{d{{x}}_i}={F}_i(\bm{x})dt+\sigma{s}(\bm{x})_{ij}d{W}_j,
\end{equation}
where $\bm{x},\bm{F} \in \mathbb{R}^{N}$, $d{W}_j$ is the increment of an $N-$dimensional brownian motion, $C_{ij}(\bm{x})={s}_{ik}(\bm{x}){s}_{jk}(\bm{x})$ is the noise covariance matrix with ${s}_{ij}(\bm{x}) \in \mathbb{R}^{N\times N}$, and $\sigma\geq 0 $. We assume that $\dot{{x}}_i(t)={F}_i(\bm{x(t)})$ has multiple steady states, so that the phase space is partitioned between the basins of attraction $B_j$ of the attractors $\Omega_j$ and the boundaries $\partial B_l$, $l=1,\ldots,L$ separating such basins.  {\color{black}Orbits initialized on the basin boundaries $\partial B_l$, $l=1,\ldots,L$ are attracted towards invariant saddles. 
Such saddles 
 $\Pi_l$, $l=1,\ldots,L$} are called edge states \cite{Grebogi1983,Robert2000,Ott2002,Vollmer2009} and can  feature chaotic dynamics  \cite{LucariniBodai2017}. In this latter case we refer to the edge states  as Melancholia states \cite{LucariniBodai2017,Lucarini2019,LucariniBodai2019arxiv}.  In absence of noise, the asymptotic state is uniquely determined by the initial condition, while noise allows trajectories to hop across boundaries between the various basins of attraction.  
\subsection{Computing the Quasi-Potential}
In the case of elliptic (and possibly hypoelliptic) diffusion processes, the Freidlin-Wentzell~\cite{Freidlin1984} theory and modifications thereof \cite{Graham1991,Hamm1994,LT:2011} show {\color{black}that in the weak-noise limit $\sigma\rightarrow 0 $ the (unique) invariant measure  can be written}  as a large deviation law:
\begin{equation}\label{eq:stationary_distr}
  \Pi_\sigma(\bm{x}) \sim \exp\left(-\frac{2\Phi(\bm{x})}{\sigma^2}\right),
\end{equation}
where the rate function $\Phi(\bm{x})$ is referred to as quasi-potential, and we have neglected the pre-exponential term. Specifically, the symbol $\sim$ in Eq. \ref{eq:stationary_distr} implies that $\Phi(\bm{x})=-2\lim_{\sigma\rightarrow 0} \sigma^2  \log \Pi_\sigma(\bm{x})$.  The function $\Phi(\bm{x})$ can be obtained as follows. We solve the stationary Fokker-Planck equation corresponding to Eq. \ref{eqapp}:
\begin{equation}
\partial_j J_j(\bm{x})=0 \quad  J_j(\bm{x}) = -{F}_j(\bm{x})  \Pi_\sigma(\bm{x}) +\sigma^2\partial_i  \left(C_{ij}(\bm{x}) \Pi_\sigma(\bm{x}) \right)
\end{equation}
where $\bm{J}$ is the current density, consider the weak noise limit, and use as ansatz the expression given in Eq. \ref{eq:stationary_distr}. We obtain the following Hamilton-Jacobi equation \cite{Gaspard2002}:
\begin{equation}\label{eq:HJE}
{F}_i(\bm{x}) \partial_i \Phi(\bm{x})+C_{ij}(\bm{x})  \partial_i \Phi(\bm{x}) \partial_j \Phi(\bm{x}) =0.
\end{equation}
The previous equation allows one to express $\Phi$ in terms of the drift and volatility fields. The quasi-potential $\Phi$ can also be computed by solving the variational problem associated with the Freidlin-Wentzell action \cite{Nardini2016}. Finally, alternative routes for computing $\Phi$ have been proposed in \cite{Ao2004,Yin2006}. \footnote{The function $\Phi(\bm{x})$ features, in general, discontinuities in its first derivatives \cite{Graham1986}.}  
The explicit computation of $\Phi$ is, in general, far from trivial, yet of great interest in many applications; see e.g., \cite{Zhou2012} for the case of biological systems. Brackston \textit{et al.} \cite{Brackston2018} have recently proposed an algorithm for estimating $\Phi$ in the case the governing
equations are polynomial and involves solving an optimization over the coefficients of a polynomial  function. Instead, \citet{Tang2017} proposed a variational method for estimating in the the populations corresponding to each determininistic attractor without resorting to computing the invariant measure.

 Following \cite{Graham1991,Hamm1994}, we now describe the dynamical meaning of $\Phi$. Indeed, solving the previous Hamilton-Jacobi equation corresponds to the fact that it is possible to write the drift vector field as the sum of two vector fields:
\begin{equation}\label{eq:decomposition}
{F}_i(\bm{x}) = {R}_i(\bm{x})- C_{ij}(\bm{x})\partial_j \Phi(\bm{x})
\end{equation}
that are mutually orthogonal, so that  ${R}_i(\bm{x})\partial_i \Phi(\bm{x})=0$. In the case Eq  \ref{eqapp} describes a thermodynamical system near equilibrium, $\bm{R}$ defines the time reversible dynamics, while $\bm{F}-\bm{R}$ defines the irreversible, dissipative dynamics \cite{Graham1987}.  
%
%
One finds that  
\begin{equation}\label{eq:Lyap}
d\Phi(\bm{x})/dt = -C_{ij}(\bm{x})  \partial_i \Phi(\bm{x}) \partial_j \Phi(\bm{x}) +  {R}_i(\bm{x(t)})  \partial_i  \Phi(\bm{x}) = - C_{ij}(\bm{x})  \partial_i \Phi(\bm{x}) \partial_j \Phi(\bm{x}).
\end{equation}
As a result, just as in the case of gradient flow, $\Phi$ has the role of a Lyapunov function whose decrease describes the convergence on an orbit to the attractor. Specifically, $\Phi(\bm{x})$ has local minima at the deterministic attractors $\Omega_j$, $j=1,\ldots, J$ and has a saddle behaviour at the edge states  $\Pi_l$, $l=1,\ldots,L$. If an attractor (edge state) is chaotic, $\Phi$ has constant value over its support, which can then be a strange set \cite{Graham1991,Hamm1994}. The standard case of gradient flow and noise correlation matrix proportional to the identity (obtained by setting ${F}_i(\bm{x})=-\partial_i U(\bm{x})$ and $C_{ij}(\bm{x})=\bm{1}$) is immediately recovered as case where $\Phi=U$. In this case, $\dot U(\bm{x})=-\partial_i U(\bm{x}) \partial_i U(\bm{x}) <0$ and  $U(\bm{x})$ is a Lyapunov function.

\subsection{Noise-induced Escape from the Attractor} 
Moreover, the probability that an orbit with initial condition in $B_j$ does not escape from it over a time $p$ decays as:
\begin{equation}\label{eq:tt_distr}
 P_{j,\sigma}(p) =\bar{r}_{j,\sigma}\exp\left(-\bar{r}_{j,\sigma}{p}\right), \quad \bar{r}_{j,\sigma}\sim \exp\left(-\frac{2\Delta\Phi_j}{\sigma^2}\right)
\end{equation}
where $\bar{r}_{j,\sigma}$ is the expected escape time and 
%
where $\Delta\Phi_j=\Phi(\Pi_l)-\Phi(\Omega_j)$ is the lowest {\color{black}pseudo-}potential barrier height \cite{LT:2011}, i.e. $\Phi$ has the lowest value in $\Pi_l$ compared to all the other edge states neighbouring $\Omega_j$. In general, one may need to add a correcting prefactor to $P_{j,\sigma}(p)$. \cite{LT:2011}. Equation \ref{eq:tt_distr} defines the residence-time distribution for basin of attraction $B_j$. See Appendix \ref{residence} for further discussion on this key statistical property.

Note that $\bar{r}_{j,\sigma}$ given in Eq. \ref{eq:tt_distr} does not contain the pre-exponential factor, as opposed to Eq. \ref{kramers}. \citet{Bouchet2016} provided an expression for such pre-exponential factor for general non-equilibrium diffusion processes under the assumption that attractors and edge states are simple points, thus generalising the results by \citet{Bovier2004}. As we aim at treating also a more general setting for the geometry of attractors and edge states, we pay below the price of having to take the pre-exponential factors as phenomenological parameters one can find from experiments or numerical simulations. We also remark that, in the zero-noise limit, the transition paths outside a basin of attraction follow the instantons. Instantons are defined as solutions of 
\begin{equation}\label{eqapp2}
d{{{x}}_i}/dt={\tilde{F}}_i(\bm{x})={R}_i(\bm{x})+C_{ij}(\bm{x})\partial_j \Phi(\bm{x})
\end{equation}
that connect a point in $\Omega_j$ to a point in $\Pi_l$. Instantonic trajectories have a reversed component of the gradient contribution to the vector field compared to regular - relaxation - trajectories.

\section{Stochastic Resonance}\label{StoRes}
Let's now assume that, generalising Eq. \ref{eq1}, we perturb the Eq. \ref{eqapp} as follows:
\begin{equation}\label{eqappp}
{d{{x}}_i}={F}_i(\bm{x})dt+\epsilon G_i(\bm{x})+\sigma{s}_{ij}(\bm{x})d{W}_j,
\end{equation}
where $\epsilon$ is a small parameter. 
As a result of the perturbation, the rate function $\Phi_\epsilon(\bm{x})$ will depend on the parameter $\epsilon$. Assuming $\epsilon$ small, in the spirit of linear response theory, we can expand $\Phi_\epsilon(\bm{x})=\Phi(\bm{x})+\epsilon \Psi(\bm{x})+h.o.t.$. In the standard one-dimensional case described by Eq. \ref{eq1}, one has $\Psi({x})=x.$ Substituting this expansion in Eq. \ref{eq:HJE} and collecting the first order terms, we obtain:
\begin{eqnarray}\label{eq:HJEpert}
&(F_i(\bm{x})) + 2C_{ij}(\bm{x})\partial_j \Phi(\bm{x}))  \partial_i \Psi(\bm{x})= - G_i(\bm{x}) \partial_i \Phi(\bm{x}). 
\end{eqnarray}
Solving the previous linear equation with respect to $\Psi(\bm{x})$ allows us to derive the first order correction to the rate function, so that $\Phi\rightarrow \Phi+\epsilon \Psi$, with the ensuing modifications in, e.g, Eqs.  \ref{eq:stationary_distr} and \ref{eq:tt_distr}.  In the latter, taking again a linear approximation, the quasi-potential difference is evaluated by considering the unperturbed attractor and edge state. \citet{Nardini2016} showed in great generality that the correction term $\Psi$ can indeed be found and proposed an algorithmic procedure to compute the perturbative terms at all orders in $\epsilon$. 

We now consider the simple case of a bistable system, such that, in the weak-noise limit\footnote{We do not take the limit $\sigma\rightarrow 0$ because this leads to concentrating the measure over the deterministic attractor featuring the lower value of the quasi-potential, leading to the disappearance of bistability; see \cite{LucariniBodai2019arxiv} for a discussion of an associated first order phase transition in a climate model.}, the escape rate (inverse of the expected escape time) from the basins of attraction $B_j$ is:
\begin{align}
{r_{j,\sigma,\epsilon}}&= A_j \exp\left(-\frac{2\Delta\Phi_j+2\epsilon\Delta\Psi_j}{\sigma^2}\right)=A_j \exp\left(-\frac{2\Delta\Phi_j}{\sigma^2}\right)\exp\left(-\frac{2\epsilon\Delta\Psi_j}{\sigma^2}\right)\label{eq:escape}\\
&\approx A_j \exp\left(-\frac{2\Delta\Phi_j}{\sigma^2}\right)\left(1-2\epsilon \frac{\Delta \Psi_1}{\sigma^2}\right)+o(\epsilon^2)\label{pert}\\
&=r_{j,\sigma}-\epsilon \alpha_{j,\sigma}+o(\epsilon^2)  \qquad \alpha_{j,\sigma}= 2 \frac{\Delta \Psi_j}{\sigma^2} r_{j,\sigma}\label{pert2}
\end{align}
where in the last passage we have assumed $\alpha_j/r_j \ll 1$, and $h.o.t.$ indicates higher order terms. We have explicit expressions for the rate in terms of the quasi-potential of the system. The escape times implied by the rates in Eq. \ref{eq:escape}-\ref{pert2} are very long compared to the dynamical time scales of the system within basins of attraction. In our case, the prefactors $A_j$, $j=1,2$ 
should be estimated from numerical experiments performed with different values of the noise strength $\sigma$. Nonetheless, unless $A_1/A_2$ is very different from 1 (which amounts to having a radically different properties of the quasi-potential near the two attractors), the results below depend  weakly (compared to $\sigma$) on $A_1/A_2$. 

We now treat the case of a time-dependent variant of the perturbed evolution given in Eq. \ref{eqappp}, where we consider $\epsilon\rightarrow \epsilon \cos(\omega t)$. If the period of the oscillation $T=2\pi/\omega$ is  much longer than the internal time scales of the system within each attractor, we obtain that, using a quasi-adiabatic approximation \cite{Gammaitoni1998}, the escape rates of the perturbed system can be written as:
\begin{equation}\label{eq:escapepert2}
r_{j,\sigma,\epsilon}(t) =  r_{j,\sigma}-\epsilon \alpha_{j,\sigma} \cos(\omega t)+o(\epsilon^2).
\end{equation}
We then perform a coarse graining and consider the two-state system  corresponding to the two unperturbed attractors $\Omega_1 $ and $\Omega_2$. The master equation for the population of state 1, $n_1(t)$, is:
\begin{equation}\label{eq:population}
\dot n_1(t)=r_2(t)-(r_1(t)+r_2(t))n_1(t)\\
\end{equation}
where $n_1(t)+n_2(t)=1$. %

The construction of a meaningful master equation relies on the presence of clear time-scale separation
between the relaxation motions near each attractor and those across the edge state, which depends critically on the presence of weak noise.

Within the two-state approximation, the time-dependent expectation value of given observable $O(\bm{x})$ is $\langle O\rangle (t) = n_1(t)\langle O \rangle_1 +  n_2(t)\langle O \rangle_2$, where $\langle O\rangle_j$ is the expectation value of $O$ in the measure supported on the unperturbed attractor $\Omega_j$. In the usual case described by Eq. \ref{eq1}, one typically choses $O=x$.

In the limit of weak forcing, the asymptotic oscillatory behaviour of  $n_1$, realised after transients have died out (this happens over a time scale $\tau=1/(r_{1,\sigma}+r_{2,\sigma})$) can be found by proposing the ansatz solution $n_1(t)=c+\epsilon R \cos(\omega t- \phi)$ in Eq. \ref{eq:population} and keeping the terms proportional to $\epsilon^0$ and $\epsilon^1$. One finds:
\begin{equation}
c=\frac{r_{2,\sigma}}{r_{1,\sigma}+r_{2,\sigma}}=\frac{1}{1+\frac{A_1}{A_2} \exp\left(-2\frac{\Delta\Phi_1-\Delta\Phi_2}{\sigma^2}\right)},
\end{equation}
\begin{align}
R&=\frac{|\alpha_{2,\sigma}r_{1,\sigma}-\alpha_{1,\sigma}r_{2,\sigma}|}{(r_{1,\sigma}+r_{2,\sigma})\left(\omega^2+(r_{1,\sigma}+r_{2,\sigma})^2\right)^{1/2}}\\
&=2\frac{|\Psi_1-\Psi_2|}{\sigma^2}\frac{r_{1,\sigma}r_{2,\sigma}}{(r_{1,\sigma}+r_{2,\sigma})\left(\omega^2+(r_{1,\sigma}+r_{2,\sigma})^2\right)^{1/2}}\\
&=\frac{2|\Psi_1- \Psi_2|\prod_{j=1}^2 A_j \exp\left(\frac{-2\Delta\Phi_j}{\sigma^2}\right)
}{\sigma^2\sum_{j=1}^2 A_j \exp\left(\frac{-2\Delta\Phi_j}{\sigma^2}\right)
\left(\omega^2+(\sum_{j=1}^2 A_j \exp\left(\frac{-2\Delta\Phi_j}{\sigma^2}\right)
)^2\right)^{1/2}}\label{erre},
\end{align}
\begin{equation}
\phi=\arctan\left(\frac{\omega}{r_{1,\sigma}+r_{2,\sigma}}\right)=\arctan\left(\frac{\omega}{ \sum_{j=1}^2 A_j \exp\left(\frac{-2\Delta\Phi_j}{\sigma^2}\right)}\right).
\end{equation}
The constant $c$ gives the unperturbed result one obtains by setting $\epsilon=0$,  while the phase difference between forcing and response is given by $\phi$.

The value of $R$ indicates whether we are in SR conditions or not, because $R$ measures the strength of the periodic motion of the ensemble mean of the trajectories. $R$ tends to zero as $\sigma\rightarrow 0$ and $\sigma\rightarrow\infty$ (keeping in mind that the latter limit goes against our weak noise assumption), and one expects that a maximum for  $R$ is achieved for intermediate values of $\sigma$. Such maximum defines conditions of SR. As discussed in \cite{Bouzat1999,Qiao2016}, the resonance is, \textit{ceteris paribus}, weakened by the presence of strong asymmetries in the system. Taking the standard symmetric case where $A_1=A_2$ and $\Delta \Phi_1=\Delta \Phi_2$, one gets, by maximizing $R$, the following trascendental equation for $\sigma$ defining the SR condition:
\begin{equation}
4r^2_{1,\sigma}(SR)=4A_1^2\exp(-4\Delta\Phi_1/\sigma_{SR}^2)=\omega^2\left(2\Delta \Phi_1/\sigma_{SR}^2-1\right).
\end{equation}
The resonance condition we find agrees, obviously, with the result presented in \cite{Gammaitoni1998}; the main improvement we get in our result is that we can relate all parameters in the previous equation to the unperturbed equations of motion via $\Phi$. We will comment below on the relevance of the specific functional form of the perturbation field $\bm{G}$. Note that, by definition, $\Psi_1- \Psi_2=\Delta \Psi_2-\Delta \Psi_1$. 

A second measure of SR\footnote{Different measures of the quality of the SR based on the synchronization between the periodic forcing and occurrence of transitions between the two basins of attraction have been proposed in \cite{Herrmann2005}.} is obtained by studying under which conditions the periodic forcing and the noise  interact constructively to create in the power spectrum of a general observable a strong spectral feature at the frequency $\omega$ of the periodic forcing. We study the $t-$averaged correlation function for a general observable $O$:
\begin{equation}
C_O(\tau)=\left\langle \lim_{t_0\rightarrow -\infty}\langle O(t+\tau)O(\tau)|O(t_0)t_0\rangle\right\rangle_t
\end{equation}
and, in particular of its symmetrized Fourier Transform $S^s_O(\nu)=S_O(\nu)+S_O(-\nu)$, where $S_O(\nu)=\mathcal{F}\{C_O(\tau)\}$ is the Fourier Transform of $C_O(\tau)$ and $\nu$ is the angular frequency. In order to find the correct expression of $S^s_O(\nu)$ one needs to consider transient behaviour as well, as opposed to the case of the estimate of $R$ above.  Following the careful calculations in \cite{Bouzat1999}, one finds that:
\begin{equation}
S^s_O(\nu)=S^s_{sing}(\nu)+S^s_{cont}(\nu)=4\pi S_0 \delta(\nu)+2\pi \epsilon^2S_2\delta(\nu-\omega)+\epsilon^2 \Sigma(\nu)+4S_1\frac{r_{1,\sigma}+r_{2,\sigma}}{(r_{1,\sigma}+r_{2,\sigma})^2+\nu^2}
\end{equation}
where the first two terms refer to the singular components of the spectrum and the second two describe the continuum component. Specifically, one has:
\begin{equation}
S_0=\frac{\left(r_{2,\sigma}\langle O \rangle_1 -r_{1,\sigma}\langle O \rangle_2 \right)^2 }{\left(r_{1,\sigma}+r_{2,\sigma}\right)^2}
\end{equation}
\begin{equation}
S_2=2\frac{\left(\langle O \rangle_1 -\langle O \rangle_2 \right)^2 |\Psi_1-\Psi_2|^2 r_{1,\sigma}^2r_{2,\sigma}^2}{\sigma^4\left(r_{1,\sigma}+r_{2,\sigma}\right)^2((r_{1,\sigma}+r_{2,\sigma})^2+\omega^2)}
\end{equation}
\begin{equation}
S_1= \frac{\left(\langle O \rangle_1 -\langle O \rangle_2 \right)^2r_{1,\sigma}r_{2,\sigma}}{(r_{1,\sigma}+r_{2,\sigma})^2}
\end{equation}
while the rather convoluted (smooth) function $\Sigma$ is not reported here for reasons that become apparent below. Note that the zero-frequency component can be removed by redefining $O\rightarrow O-\left(r_{2,\sigma}\langle O \rangle_1 -r_{1,\sigma}\langle O \rangle_2 \right)$, i.e., by removing its unperturbed ensemble mean. Instead, all the other terms disappear if $\langle O \rangle_1= \langle O \rangle_2 $, i.e. if we choose an observable that does not distinguish between the two unperturbed attractors (e.g. choosing $O=x^2$ in the  setting  of Eq. \ref{eq1}).

Following \cite{Gammaitoni1998}, one defines the (linear) spectral amplification $SNR$ as follows:
\begin{equation}
SNR=\lim_{\epsilon\rightarrow 0}\frac{1}{\epsilon^2}\lim_{\Delta \omega\rightarrow 0}\frac{\int_{\omega-\Delta \omega}^{\omega+\Delta\omega}d\nu S_{sing}(\nu)}{S_{cont}(\omega)}=\frac{\pi}{2}\frac{S_2}{S_1}\frac{\left(r_{1,\sigma}+r_{2,\sigma}\right)^2+\omega^2}{r_{1,\sigma}+r_{2,\sigma}}
\end{equation}
which leads to the final result:
\begin{equation}
SNR=\pi\frac{|\Psi_1-\Psi_2|^2r_1r_2}{\sigma^4(r_{1,\sigma}+r_{2,\sigma})}=\pi\frac{|\Psi_1-\Psi_2| ^2}{\sigma^4}\frac{\prod_{j=1,2}A_j\exp(-2\Delta\Phi_j/\sigma^2)}{\sum_{j=1,2}A_j\exp(-2\Delta\Phi_j/\sigma^2)}.\label{SR}
\end{equation}
As well know, in the small $\epsilon$ limit, $SNR$ does not depend on $\omega$. Additionally, the parameter $SNR$ is clearly maximized in the symmetric case $A_1=A_2$, $\Delta\Phi_1=\Delta\Phi_2$, where we obtain $SNR=\pi/2 |\Psi_1-\Psi_2| ^2 A_1\exp(-2\Delta \Phi_1/\sigma^2)/\sigma^4$. In the symmetric case, $SNR$ is maximized if $\sigma^2=\Delta \Phi_1$, regardless of the perturbation field, which generalises what given in \cite{Gammaitoni1998}.

The expression of $R$ in Eq. \ref{erre} and of $SNR$ in Eq. \ref{SR} indicate that, in the weak-perturbation and weak noise limit the choice of the perturbation field $\bm{G}$ impacts the strength of the  signal (both in conditions of SR or not)  exclusively through the factor $|\Psi_1-\Psi_2|$. Clearly, perturbation fields $\bm{G}$'s differing in the  transversal (with respect to the gradient structure, see Eq. \ref{eq:decomposition}) component have the same effect in terms of SR. 

In particular, we find that SR is entirely suppressed if $\Psi_1=\Psi_2$. In other terms, if the change in the quasi-potential is the same in $\Omega_1$ and $\Omega_2$, there is no SR phenomenon at all. Again, this condition can be realised for a very large class of non trivial $\bm{G}$'s. In this case, adiabatically we see an periodic increase and decrease of the both $r_{1,\sigma}$ and $r_{2,\sigma}$. This amounts to a slow modulation in the  overall inter-well time scale of the system and has no differential effects on the transition $1\rightarrow 2$ and $2\rightarrow 1$. 

\section{Conclusions}\label{conclu}
Our analysis above bridges the investigation of the statistical properties of general non-equilibrium systems with that of SR. Our findings give as special case the classical results valid for systems obeying detailed balance, as in the case of N-dimensional gradient flows forced by standard additive noise. 

Our approach revolves around the computation of the  quasi-potential $\Phi$ of the unperturbed dynamics.  
The quasi-potential is proportional to the rate function defining the large deviation law describing the invariant measure of the system, and controls the rate of escape of trajectories from a given basin of attraction through a relevant edge state. Additionally, the quasi-potential is associated with the decomposition of the drift term into the sum of a gradient component  (defined by the quasi-potential) and a transversal component, which are mutually orthogonal. Finally, $\Phi$  acts as a Lyapunov function for the dynamics, and reaches local minima at the attractor(s) of the system. In the case of multistable system, $\Phi$ has a saddle behaviour at the edge states embedded in the boundaries of the basins of attraction.  

If we consider a system possessing two metastable states undergoing a periodic modulation to the drift term, the previous framework allows to derive some of the main classical results of SR within the two-state approximation with the great advantage that the parameters contained in the SR conditions can be derived from the drift and volatility terms of rather general equations of motion. Indeed, we need to compute the correction $\Psi$ to the quasi-potential $\Phi$, the former being associated with the spatial pattern of the periodically driven perturbation field. We are able to deal with the fairly general case where the the attractors and/or the edge state feature chaotic dynamics.

We have clarified that SR is an intrinsic property of the unperturbed system, which is catalysed by the presence of the periodically oscillating perturbation field. The details of the perturbation to the drift term impact the SR intensity through a simple factor depending on $\Psi$. Additionally, we have explained that one can define classes of equivalence of perturbations in terms of their SR properties, with each class constituted by elements differing only by the transversal component with respect to the gradient structure determined by $\Phi$. Similarly, one finds that - see Appendix \ref{residence} - the statistical properties of  the residence times can be fully described in terms of $\Phi$ and $\Psi$.

Our results can in principle be extended to the case of multiple metastable states connected through a potentially non-trivial network of channels defined by the edge states between them. Addressing the challenges of a complex topology of transition paths would probably benefit from taking advantage from the sophisticated super-symmetric techniques developed by \citet{Tanase-Nicola2004} in the context of noisy gradient flows.

A relevant improvement to our results would come from the possibility of using a general formula for the pre-exponential factor of the escape rates valid also for the case of nontrivial attracting and saddle sets. The lack of such a  formula makes our finding somewhat phenomenological, yet hopefully useful. Clearly, the results above would greatly benefit from a more rigorous level of mathematical formulation, which is currently beyond the abilities of the author. 

The findings above pave the way for the rigorous study of SR-related phenomena in many systems, especially taking into account that the assumption of a (one-dimensional) gradient structure for the drift component of the flow is far from verified (or meaningful, in fact) in general; see discussion in \cite{LucariniFarandaWilleit2012}.  
We foresee applications in the many areas where SR has proved to be a valuable and useful concept. The fact that several algorithms for computing the quasi-potential have been recently presented in the literature makes possible a systematic exploitation of the results discussed here.

As for the inclination of the author, further investigation will focus on the study of SR in geophysical systems, where multistability is often encountered and of great relevance \cite{Lenton2008,Ghil2019}. In the spirit of the first investigations of SR, the author will attempt a careful numerical and analytical examination of SR in the multistable climate model presented in \cite{Lucarini2019,LucariniBodai2019arxiv}, where noise permits transitions between the competing snowball and warm  climate states. Furthermore, following \cite{Alley2001,Ganopolski2002}, the author aims at studying using the formalism presented here the SR acting between the two competing \textit{on} and \textit{off} states of the deep ocean circulation, featuring a strong and virtually absent overturning circulation in the Atlantic ocean, respectively \cite{Rahmstorf2005,Kuhlbrodt2007}. 

\section*{Acknowledgments}
The author wish to thank T. Bodai, F. Bouchet, F. Flandoli, X.-M. Li, T. Lelievre, T. Kuna, J. Kurchan, G. Pavliotis,  I. Pavlyukevich, and T. Tel for useful exchanges, and  acknowledges the support received from the EU Horizon2020 projects Blue-Action (Grant No. 727852) and TiPES (Grant No. 820970), and from the DFG SFB/Transregio Project TRR 181. This paper has been written while the author was participating to the scientific programme \textit{The mathematics of climate and the environment} (September-December 2019) held at the I. H. Poincar\'e in Paris. 
\appendix
\section{Residence-Time Distribution}\label{residence}
The residence time for the system in a state is the random variable describing the time interval spent consecutively in such a state before a noise-induced transition occurs, and the system reaches another state. The statistics of the residence times is one of the essential properties of stochastically perturbed multistable systems, and the typical exponential residence-time distribution for the system in absence of periodic forcing has been introduced in Eq. \ref{eq:tt_distr}. When a periodic forcing is applied, the simple exponential law is fundamentally altered, in such a way that escape from a state is periodically enhanced and suppressed. 

In the classic setting of Eq. \ref{eq1}, one finds peaks in the residence time in each state (modulated by the exponential decay associated to the time-averaged transition rate) at times $p=(2n+1) T/2$, as a result of the alternating increase and decrease of the potential barrier for either state. Key is the fact that the time-dependent anomaly of the potential is at all times in opposite phase in the two states. If the system is in state 1 at time zero, after having performed a transition from state 2 when the potential barrier of state 2 is at a minimum, it has to wait half a period before the potential barrier of state 1 reaches a minimum. This explains the first peak. If the transition does not take place, the system has to wait a full period before reaching again favourable conditions for the jump. The intensity of such an effect depends dramatically on $\epsilon/\sigma^2$, up to reaching an almost perfect synchronization of the transitions with the phase of the periodic forcing \cite{Gammaitoni1998}. A very accurate analysis of the residence-time statistics has been given by \citet{Choi1998} and \citet{Berglund2005} in the case of symmetric potential described by Eq. \ref{eq1}. 

In what follows, we build on previous results by \citet{Coppersmith1994}, who studied residence-time statistics within the two-state approximation, and considered asymmetries in the two states and in their response to the periodic perturbation. We rewrite the escape rate from the state $j=1,2$ as: 
\begin{equation}
r_{j,\sigma,\epsilon}(t)=\overline{r_{j,\sigma,\epsilon}}+\delta\{r_{j,\sigma,\epsilon}\}(t)\label{eqa1}
\end{equation}
where $\overline{r_{j,\sigma,\epsilon}}=1/T\int_t^{T+t} dt_1 r_{j,\sigma,\epsilon}(t_1)$ and $\int_t^{T+t} dt_1 \delta\{r_{j,\sigma,\epsilon}(t_1)\}=0$, with $T=2\pi/\omega$. We now consider Eq. \ref{eq:escape} and substitute $\epsilon\rightarrow\epsilon\cos(\omega t)$. By expanding in power series the exponential function, one obtains:
\begin{equation}
\overline{r_{j,\sigma,\epsilon}}=A_{j}\exp\left(-\frac{2\Phi_j}{\sigma^2}\right)\mathcal{F}\left(-\frac{2\epsilon\Phi_j}{\sigma^2}\right)=r_{j,\sigma}\mathcal{F}\left(-\frac{2\epsilon\Phi_j}{\sigma^2}\right)\label{eqa2}
\end{equation}
where $\mathcal{F}(x)=\sum_{j=0}^\infty  \frac{1}{2^{2j}}\frac{1}{(j!)^2}x^{2j}$. The function $\mathcal{F}$ increases rapidly with $x$, with $\mathcal{F}(0)=1$ (see the linear expansion of $r_{j,\sigma,\epsilon}(t)$ in Eq. \ref{eq:escapepert2}), while the first non-trivial term is given by $1/4x^2$,  Therefore, as a result of the periodic modulation of the argument of the exponent, the average rate of escape within a period is larger than the unperturbed one, with the first correction being quadratic in $\epsilon$. We then derive 
\begin{align}
\delta\{r_{j,\sigma,\epsilon}\}(t)&=A_j \exp\left(-\frac{2\Delta\Phi_j}{\sigma^2}\right)\exp\left(-\frac{2\epsilon\cos(\omega t)\Delta\Psi_j}{\sigma^2}\right)-A_{j}\exp\left(-\frac{2\Delta\Phi_j}{\sigma^2}\right)\mathcal{F}\left(-\frac{2\epsilon\Phi_j}{\sigma^2}\right)\label{eqa3}\\
&=r_{j,\sigma}\left(\exp\left(-\frac{2\epsilon\cos(\omega t)\Delta\Psi_j}{\sigma^2}\right)-\mathcal{F}\left(-\frac{2\epsilon\Phi_j}{\sigma^2}\right)\right)\label{eqa4}
\end{align}
Following \citet{Coppersmith1994}, we have that the residence-time distribution $P_{j,\sigma,\epsilon}(p)$, $j=1,2$ can be written as:
\begin{equation}
P_{j,\sigma,\epsilon}(p)=N_{j,\sigma,\epsilon}\exp(-\overline{r_{j,\sigma,\epsilon}}p)G_{j,\sigma,\epsilon}(p)\label{eqa5}
\end{equation}
where $G_{j,\sigma,\epsilon}(p)$ is periodic of period $T$, i.e. $G_{j,\epsilon}(p+T)=G_{j,\epsilon}(p)$  and $N_{j,\sigma,\epsilon}$ is a normalization constant such that $\int_0^\infty d\tau P_{j,\sigma,\epsilon}(\tau)=1$. The function $G_{j,\sigma,\epsilon}(p)$ is:
\begin{equation}
G_{j,\sigma,\epsilon}(p)=\int_0^{T} dt_1 r_{j,\sigma,\epsilon}(t_1+p) \exp\left(-\int_{t_1}^{t_1+\tau} dt_2 \delta\{r_{j,\sigma,\epsilon}\}(t_2) \right) Y_{j,\sigma,\epsilon}(t_1)
\end{equation}
where the function $Y_{j,\sigma,\epsilon}$ is defined as follows:
\begin{equation}
Y_{j+1,\sigma,\epsilon}(s)=\frac{r_{j,\sigma,\epsilon}(s)}{1-\exp(\overline{r_{j,\sigma,\epsilon}}T)}
\int_0^{T} dt_1 Y_{j,\sigma,\epsilon}(s-t_1)\exp(\overline{r_{j,\sigma,\epsilon}}t_1) \exp\left(-\int_{s-t_1}^{s} dt_2 \delta\{r_{j,\sigma,\epsilon}\}(t_2) \right)
\end{equation}
where $Y_{j+1,\sigma,\epsilon}(s)=Y_{1,\sigma,\epsilon}(s)$ if $j=2$. The previous equation provides a link between the escape rates from the two states and can be solved recursively starting, e.g., from the initial ansatz $Y_{1,\sigma,\epsilon}(s)=const$. A closed expression for $Y_{1,\sigma,\epsilon}$ and $Y_{2,\sigma,\epsilon}$ can be found by assuming $\epsilon$ small and taking a linear approximation. If $\epsilon=0$, one recovers the result shown in Eq. \ref{eq:tt_distr}, where $G_{j,\sigma,\epsilon}(s)|_{\epsilon=0}=r_{j,\sigma}$, $Y_{j+1,\sigma,\epsilon}(s)|_{\epsilon=0}=1/T$, and $N_{j,\sigma,\epsilon}|_{\epsilon=0}=1$, $j=1,2$. Using Eqs. \ref{eqa1}-\ref{eqa4}, one can define the residence-time distribution in terms of the drift and volatility field of the system, according to the procedure described in Sect. \ref{mathframe}, and exclusively through the definition of the quasi-potentials $\Phi$ and $\Psi$; the transversal components of the fields do not matter. This is a key advantage of the formulation of the problem proposed in this paper. 

Building on \cite{Coppersmith1994}, one understands that, depending on the intensity of the perturbation quasi-potential $\Psi$ and of its asymmetry in the two states 1 and 2, and on the asymmetry of the unperturbed quasi-potential $\Phi$ between the two states, the intensity of the periodic modulation contained in the function $G_{j,\sigma,\epsilon}(p)$ can vary enormously.  In particular, it is reasonable to guess that  the properties of the periodic modulation of $G_{j,\sigma,\epsilon}(p)$ are strongly influenced by the presence of a difference in the perturbation of the quasi-potential $\Psi_1$ and $\Psi_2$ at all times (the classical setting corresponds to $\Delta\Psi_1=-\Delta \Psi_2$). One then expects that the presence of characteristic peaks at times $p=(2n+1) T/2$ mentioned above depends critically on the quantity  $|\Psi_1- \Psi_2|/\sigma^2$. Nonetheless, the investigation of these properties and the analysis of the conditions conducive to SR from this point of view is beyond the scopes of this paper.


\begin{thebibliography}{74}%
\makeatletter
\providecommand \@ifxundefined [1]{%
 \@ifx{#1\undefined}
}%
\providecommand \@ifnum [1]{%
 \ifnum #1\expandafter \@firstoftwo
 \else \expandafter \@secondoftwo
 \fi
}%
\providecommand \@ifx [1]{%
 \ifx #1\expandafter \@firstoftwo
 \else \expandafter \@secondoftwo
 \fi
}%
\providecommand \natexlab [1]{#1}%
\providecommand \enquote  [1]{``#1''}%
\providecommand \bibnamefont  [1]{#1}%
\providecommand \bibfnamefont [1]{#1}%
\providecommand \citenamefont [1]{#1}%
\providecommand \href@noop [0]{\@secondoftwo}%
\providecommand \href [0]{\begingroup \@sanitize@url \@href}%
\providecommand \@href[1]{\@@startlink{#1}\@@href}%
\providecommand \@@href[1]{\endgroup#1\@@endlink}%
\providecommand \@sanitize@url [0]{\catcode `\\12\catcode `\$12\catcode
  `\&12\catcode `\#12\catcode `\^12\catcode `\_12\catcode `\%12\relax}%
\providecommand \@@startlink[1]{}%
\providecommand \@@endlink[0]{}%
\providecommand \url  [0]{\begingroup\@sanitize@url \@url }%
\providecommand \@url [1]{\endgroup\@href {#1}{\urlprefix }}%
\providecommand \urlprefix  [0]{URL }%
\providecommand \Eprint [0]{\href }%
\providecommand \doibase [0]{http://dx.doi.org/}%
\providecommand \selectlanguage [0]{\@gobble}%
\providecommand \bibinfo  [0]{\@secondoftwo}%
\providecommand \bibfield  [0]{\@secondoftwo}%
\providecommand \translation [1]{[#1]}%
\providecommand \BibitemOpen [0]{}%
\providecommand \bibitemStop [0]{}%
\providecommand \bibitemNoStop [0]{.\EOS\space}%
\providecommand \EOS [0]{\spacefactor3000\relax}%
\providecommand \BibitemShut  [1]{\csname bibitem#1\endcsname}%
\let\auto@bib@innerbib\@empty
\bibitem [{\citenamefont {Benzi}\ \emph {et~al.}(1981)\citenamefont {Benzi},
  \citenamefont {Sutera},\ and\ \citenamefont {Vulpiani}}]{Benzi1981}%
  \BibitemOpen
  \bibfield  {author} {\bibinfo {author} {\bibfnamefont {R.}~\bibnamefont
  {Benzi}}, \bibinfo {author} {\bibfnamefont {A.}~\bibnamefont {Sutera}}, \
  and\ \bibinfo {author} {\bibfnamefont {A.}~\bibnamefont {Vulpiani}},\ }\href
  {\doibase 10.1088/0305-4470/14/11/006} {\bibfield  {journal} {\bibinfo
  {journal} {Journal of Physics A: Mathematical and General}\ }\textbf
  {\bibinfo {volume} {14}},\ \bibinfo {pages} {L453} (\bibinfo {year}
  {1981})}\BibitemShut {NoStop}%
\bibitem [{\citenamefont {Benzi}\ \emph {et~al.}(1982)\citenamefont {Benzi},
  \citenamefont {Parisi}, \citenamefont {Sutera},\ and\ \citenamefont
  {Vulpiani}}]{Benzi1982}%
  \BibitemOpen
  \bibfield  {author} {\bibinfo {author} {\bibfnamefont {R.}~\bibnamefont
  {Benzi}}, \bibinfo {author} {\bibfnamefont {G.}~\bibnamefont {Parisi}},
  \bibinfo {author} {\bibfnamefont {A.}~\bibnamefont {Sutera}}, \ and\ \bibinfo
  {author} {\bibfnamefont {A.}~\bibnamefont {Vulpiani}},\ }\href {\doibase
  10.3402/tellusa.v34i1.10782} {\bibfield  {journal} {\bibinfo  {journal}
  {Tellus}\ }\textbf {\bibinfo {volume} {34}},\ \bibinfo {pages} {10} (\bibinfo
  {year} {1982})},\ \Eprint
  {http://arxiv.org/abs/https://doi.org/10.3402/tellusa.v34i1.10782}
  {https://doi.org/10.3402/tellusa.v34i1.10782} \BibitemShut {NoStop}%
\bibitem [{\citenamefont {{Benzi}}\ \emph {et~al.}(1983)\citenamefont
  {{Benzi}}, \citenamefont {{Parisi}}, \citenamefont {{Sutera}},\ and\
  \citenamefont {{Vulpiani}}}]{Benzi1983}%
  \BibitemOpen
  \bibfield  {author} {\bibinfo {author} {\bibfnamefont {R.}~\bibnamefont
  {{Benzi}}}, \bibinfo {author} {\bibfnamefont {G.}~\bibnamefont {{Parisi}}},
  \bibinfo {author} {\bibfnamefont {A.}~\bibnamefont {{Sutera}}}, \ and\
  \bibinfo {author} {\bibfnamefont {A.}~\bibnamefont {{Vulpiani}}},\
  }\href@noop {} {\bibfield  {journal} {\bibinfo  {journal} {{SIAM J. Appl.
  Math.}}\ }\textbf {\bibinfo {volume} {43}},\ \bibinfo {pages} {563} (\bibinfo
  {year} {1983})}\BibitemShut {NoStop}%
\bibitem [{\citenamefont {Nicolis}(1981)}]{Nicolis1981}%
  \BibitemOpen
  \bibfield  {author} {\bibinfo {author} {\bibfnamefont {C.}~\bibnamefont
  {Nicolis}},\ }in\ \href@noop {} {\emph {\bibinfo {booktitle} {Physics of
  Solar Variations}}},\ \bibinfo {editor} {edited by\ \bibinfo {editor}
  {\bibfnamefont {V.}~\bibnamefont {Domingo}}}\ (\bibinfo  {publisher}
  {Springer Netherlands},\ \bibinfo {address} {Dordrecht},\ \bibinfo {year}
  {1981})\ pp.\ \bibinfo {pages} {473--478}\BibitemShut {NoStop}%
\bibitem [{\citenamefont {Nicolis}(1982)}]{Nicolis1982}%
  \BibitemOpen
  \bibfield  {author} {\bibinfo {author} {\bibfnamefont {C.}~\bibnamefont
  {Nicolis}},\ }\href {\doibase 10.3402/tellusa.v34i1.10781} {\bibfield
  {journal} {\bibinfo  {journal} {Tellus}\ }\textbf {\bibinfo {volume} {34}},\
  \bibinfo {pages} {1} (\bibinfo {year} {1982})},\ \Eprint
  {http://arxiv.org/abs/https://doi.org/10.3402/tellusa.v34i1.10781}
  {https://doi.org/10.3402/tellusa.v34i1.10781} \BibitemShut {NoStop}%
\bibitem [{\citenamefont {McNamara}\ \emph {et~al.}(1988)\citenamefont
  {McNamara}, \citenamefont {Wiesenfeld},\ and\ \citenamefont
  {Roy}}]{McNamara1988}%
  \BibitemOpen
  \bibfield  {author} {\bibinfo {author} {\bibfnamefont {B.}~\bibnamefont
  {McNamara}}, \bibinfo {author} {\bibfnamefont {K.}~\bibnamefont
  {Wiesenfeld}}, \ and\ \bibinfo {author} {\bibfnamefont {R.}~\bibnamefont
  {Roy}},\ }\href {\doibase 10.1103/PhysRevLett.60.2626} {\bibfield  {journal}
  {\bibinfo  {journal} {Phys. Rev. Lett.}\ }\textbf {\bibinfo {volume} {60}},\
  \bibinfo {pages} {2626} (\bibinfo {year} {1988})}\BibitemShut {NoStop}%
\bibitem [{\citenamefont {Stroescu}\ \emph {et~al.}(2016)\citenamefont
  {Stroescu}, \citenamefont {Hume},\ and\ \citenamefont
  {Oberthaler}}]{Stroescu2016}%
  \BibitemOpen
  \bibfield  {author} {\bibinfo {author} {\bibfnamefont {I.}~\bibnamefont
  {Stroescu}}, \bibinfo {author} {\bibfnamefont {D.~B.}~\bibnamefont {Hume}}, \
  and\ \bibinfo {author} {\bibfnamefont {M.~K.}~\bibnamefont {Oberthaler}},\
  }\href {\doibase 10.1103/PhysRevLett.117.243005} {\bibfield  {journal}
  {\bibinfo  {journal} {Physical Review Letters}\ }\textbf {\bibinfo {volume}
  {117}},\ \bibinfo {pages} {243005} (\bibinfo {year} {2016})}\BibitemShut
  {NoStop}%
\bibitem [{\citenamefont {Wagner}\ \emph {et~al.}(2019)\citenamefont {Wagner},
  \citenamefont {Talkner}, \citenamefont {Bayer}, \citenamefont {Rugeramigabo},
  \citenamefont {H{\"a}nggi},\ and\ \citenamefont {Haug}}]{Wagner2019}%
  \BibitemOpen
  \bibfield  {author} {\bibinfo {author} {\bibfnamefont {T.}~\bibnamefont
  {Wagner}}, \bibinfo {author} {\bibfnamefont {P.}~\bibnamefont {Talkner}},
  \bibinfo {author} {\bibfnamefont {J.~C.}\ \bibnamefont {Bayer}}, \bibinfo
  {author} {\bibfnamefont {E.~P.}\ \bibnamefont {Rugeramigabo}}, \bibinfo
  {author} {\bibfnamefont {P.}~\bibnamefont {H{\"a}nggi}}, \ and\ \bibinfo
  {author} {\bibfnamefont {R.~J.}\ \bibnamefont {Haug}},\ }\href {\doibase
  10.1038/s41567-018-0412-5} {\bibfield  {journal} {\bibinfo  {journal} {Nature
  Physics}\ }\textbf {\bibinfo {volume} {15}},\ \bibinfo {pages} {330}
  (\bibinfo {year} {2019})}\BibitemShut {NoStop}%
\bibitem [{\citenamefont {Schiavoni}\ \emph {et~al.}(2002)\citenamefont
  {Schiavoni}, \citenamefont {Carminati}, \citenamefont {Sanchez-Palencia},
  \citenamefont {Renzoni},\ and\ \citenamefont {Grynberg}}]{Schiavoni2002}%
  \BibitemOpen
  \bibfield  {author} {\bibinfo {author} {\bibfnamefont {M.}~\bibnamefont
  {Schiavoni}}, \bibinfo {author} {\bibfnamefont {F.-R.}\ \bibnamefont
  {Carminati}}, \bibinfo {author} {\bibfnamefont {L.}~\bibnamefont
  {Sanchez-Palencia}}, \bibinfo {author} {\bibfnamefont {F.}~\bibnamefont
  {Renzoni}}, \ and\ \bibinfo {author} {\bibfnamefont {G.}~\bibnamefont
  {Grynberg}},\ }\href {\doibase 10.1209/epl/i2002-00134-y} {\bibfield
  {journal} {\bibinfo  {journal} {Europhysics Letters ({EPL})}\ }\textbf
  {\bibinfo {volume} {59}},\ \bibinfo {pages} {493} (\bibinfo {year}
  {2002})}\BibitemShut {NoStop}%
\bibitem [{\citenamefont {Jerome}\ and\ \citenamefont
  {Ayyagari}(2014)}]{Jerome2014}%
  \BibitemOpen
  \bibfield  {author} {\bibinfo {author} {\bibfnamefont {M.~M.}\ \bibnamefont
  {Jerome}}\ and\ \bibinfo {author} {\bibfnamefont {R.}~\bibnamefont
  {Ayyagari}},\ }\href {\doibase
  https://doi.org/10.3182/20140313-3-IN-3024.00223} {\bibfield  {journal}
  {\bibinfo  {journal} {IFAC Proceedings Volumes}\ }\textbf {\bibinfo {volume}
  {47}},\ \bibinfo {pages} {313 } (\bibinfo {year} {2014})},\ \bibinfo {note}
  {3rd International Conference on Advances in Control and Optimization of
  Dynamical Systems (2014)}\BibitemShut {NoStop}%
\bibitem [{\citenamefont {Korneta}\ \emph {et~al.}(2006)\citenamefont
  {Korneta}, \citenamefont {Gomes}, \citenamefont {Mirasso},\ and\
  \citenamefont {Toral}}]{Korneta2006}%
  \BibitemOpen
  \bibfield  {author} {\bibinfo {author} {\bibfnamefont {W.}~\bibnamefont
  {Korneta}}, \bibinfo {author} {\bibfnamefont {I.}~\bibnamefont {Gomes}},
  \bibinfo {author} {\bibfnamefont {C.~R.}\ \bibnamefont {Mirasso}}, \ and\
  \bibinfo {author} {\bibfnamefont {R.}~\bibnamefont {Toral}},\ }\href
  {\doibase https://doi.org/10.1016/j.physd.2006.05.016} {\bibfield  {journal}
  {\bibinfo  {journal} {Physica D: Nonlinear Phenomena}\ }\textbf {\bibinfo
  {volume} {219}},\ \bibinfo {pages} {93 } (\bibinfo {year}
  {2006})}\BibitemShut {NoStop}%
\bibitem [{\citenamefont {Han}\ \emph {et~al.}(2014)\citenamefont {Han},
  \citenamefont {Yang}, \citenamefont {Zeng}, \citenamefont {Wang},
  \citenamefont {Liu}, \citenamefont {Fu}, \citenamefont {Zhang},\ and\
  \citenamefont {Tian}}]{Han2014}%
  \BibitemOpen
  \bibfield  {author} {\bibinfo {author} {\bibfnamefont {Q.}~\bibnamefont
  {Han}}, \bibinfo {author} {\bibfnamefont {T.}~\bibnamefont {Yang}}, \bibinfo
  {author} {\bibfnamefont {C.}~\bibnamefont {Zeng}}, \bibinfo {author}
  {\bibfnamefont {H.}~\bibnamefont {Wang}}, \bibinfo {author} {\bibfnamefont
  {Z.}~\bibnamefont {Liu}}, \bibinfo {author} {\bibfnamefont {Y.}~\bibnamefont
  {Fu}}, \bibinfo {author} {\bibfnamefont {C.}~\bibnamefont {Zhang}}, \ and\
  \bibinfo {author} {\bibfnamefont {D.}~\bibnamefont {Tian}},\ }\href {\doibase
  https://doi.org/10.1016/j.physa.2014.04.015} {\bibfield  {journal} {\bibinfo
  {journal} {Physica A: Statistical Mechanics and its Applications}\ }\textbf
  {\bibinfo {volume} {408}},\ \bibinfo {pages} {96 } (\bibinfo {year}
  {2014})}\BibitemShut {NoStop}%
\bibitem [{\citenamefont {Alley}\ \emph {et~al.}(2001)\citenamefont {Alley},
  \citenamefont {Anandakrishnan},\ and\ \citenamefont {Jung}}]{Alley2001}%
  \BibitemOpen
  \bibfield  {author} {\bibinfo {author} {\bibfnamefont {R.~B.}\ \bibnamefont
  {Alley}}, \bibinfo {author} {\bibfnamefont {S.}~\bibnamefont
  {Anandakrishnan}}, \ and\ \bibinfo {author} {\bibfnamefont {P.}~\bibnamefont
  {Jung}},\ }\href {\doibase 10.1029/2000PA000518} {\bibfield  {journal}
  {\bibinfo  {journal} {Paleoceanography}\ }\textbf {\bibinfo {volume} {16}},\
  \bibinfo {pages} {190} (\bibinfo {year} {2001})},\ \Eprint
  {http://arxiv.org/abs/https://agupubs.onlinelibrary.wiley.com/doi/pdf/10.1029/2000PA000518}
  {https://agupubs.onlinelibrary.wiley.com/doi/pdf/10.1029/2000PA000518}
  \BibitemShut {NoStop}%
\bibitem [{\citenamefont {Ganopolski}\ and\ \citenamefont
  {Rahmstorf}(2002)}]{Ganopolski2002}%
  \BibitemOpen
  \bibfield  {author} {\bibinfo {author} {\bibfnamefont {A.}~\bibnamefont
  {Ganopolski}}\ and\ \bibinfo {author} {\bibfnamefont {S.}~\bibnamefont
  {Rahmstorf}},\ }\href {\doibase 10.1103/PhysRevLett.88.038501} {\bibfield
  {journal} {\bibinfo  {journal} {Phys. Rev. Lett.}\ }\textbf {\bibinfo
  {volume} {88}},\ \bibinfo {pages} {038501} (\bibinfo {year}
  {2002})}\BibitemShut {NoStop}%
\bibitem [{\citenamefont {H\"anggi}(2002)}]{Hanggi2002}%
  \BibitemOpen
  \bibfield  {author} {\bibinfo {author} {\bibfnamefont {P.}~\bibnamefont
  {H\"anggi}},\ }\href {\doibase
  10.1002/1439-7641(20020315)3:3<285::AID-CPHC285>3.0.CO;2-A} {\bibfield
  {journal} {\bibinfo  {journal} {ChemPhysChem}\ }\textbf {\bibinfo {volume}
  {3}},\ \bibinfo {pages} {285} (\bibinfo {year} {2002})}\BibitemShut {NoStop}%
\bibitem [{\citenamefont {Lalwani}\ and\ \citenamefont
  {Brang}(2019)}]{Lalwani2019}%
  \BibitemOpen
  \bibfield  {author} {\bibinfo {author} {\bibfnamefont {P.}~\bibnamefont
  {Lalwani}}\ and\ \bibinfo {author} {\bibfnamefont {D.}~\bibnamefont
  {Brang}},\ }\href {\doibase 10.1098/rstb.2019.0029} {\bibfield  {journal}
  {\bibinfo  {journal} {Philosophical Transactions of the Royal Society B:
  Biological Sciences}\ }\textbf {\bibinfo {volume} {374}},\ \bibinfo {pages}
  {20190029} (\bibinfo {year} {2019})},\ \Eprint
  {http://arxiv.org/abs/https://royalsocietypublishing.org/doi/pdf/10.1098/rstb.2019.0029}
  {https://royalsocietypublishing.org/doi/pdf/10.1098/rstb.2019.0029}
  \BibitemShut {NoStop}%
\bibitem [{\citenamefont {McDonnell}\ and\ \citenamefont
  {Abbott}(2009)}]{McDonnell2008}%
  \BibitemOpen
  \bibfield  {author} {\bibinfo {author} {\bibfnamefont {M.~D.}\ \bibnamefont
  {McDonnell}}\ and\ \bibinfo {author} {\bibfnamefont {D.}~\bibnamefont
  {Abbott}},\ }\href {\doibase 10.1371/journal.pcbi.1000348} {\bibfield
  {journal} {\bibinfo  {journal} {PLoS computational biology}\ }\textbf
  {\bibinfo {volume} {5}},\ \bibinfo {pages} {e1000348} (\bibinfo {year}
  {2009})}\BibitemShut {NoStop}%
\bibitem [{\citenamefont {Ward}\ \emph {et~al.}(2007)\citenamefont {Ward},
  \citenamefont {Doesburg}, \citenamefont {Kitajo}, \citenamefont {MacLean},\
  and\ \citenamefont {Roggeveen}}]{Ward2007}%
  \BibitemOpen
  \bibfield  {author} {\bibinfo {author} {\bibfnamefont {L.}~\bibnamefont
  {Ward}}, \bibinfo {author} {\bibfnamefont {S.}~\bibnamefont {Doesburg}},
  \bibinfo {author} {\bibfnamefont {K.}~\bibnamefont {Kitajo}}, \bibinfo
  {author} {\bibfnamefont {S.}~\bibnamefont {MacLean}}, \ and\ \bibinfo
  {author} {\bibfnamefont {A.}~\bibnamefont {Roggeveen}},\ }\href {\doibase
  10.1037/cjep2006029} {\bibfield  {journal} {\bibinfo  {journal} {Canadian
  journal of experimental psychology = Revue canadienne de psychologie
  expérimentale}\ }\textbf {\bibinfo {volume} {60}},\ \bibinfo {pages} {319}
  (\bibinfo {year} {2007})}\BibitemShut {NoStop}%
\bibitem [{\citenamefont {Gammaitoni}\ \emph {et~al.}(1998)\citenamefont
  {Gammaitoni}, \citenamefont {H\"anggi}, \citenamefont {Jung},\ and\
  \citenamefont {Marchesoni}}]{Gammaitoni1998}%
  \BibitemOpen
  \bibfield  {author} {\bibinfo {author} {\bibfnamefont {L.}~\bibnamefont
  {Gammaitoni}}, \bibinfo {author} {\bibfnamefont {P.}~\bibnamefont
  {H\"anggi}}, \bibinfo {author} {\bibfnamefont {P.}~\bibnamefont {Jung}}, \
  and\ \bibinfo {author} {\bibfnamefont {F.}~\bibnamefont {Marchesoni}},\
  }\href {\doibase 10.1103/RevModPhys.70.223} {\bibfield  {journal} {\bibinfo
  {journal} {Rev. Mod. Phys.}\ }\textbf {\bibinfo {volume} {70}},\ \bibinfo
  {pages} {223} (\bibinfo {year} {1998})}\BibitemShut {NoStop}%
\bibitem [{\citenamefont {Anishchenko}\ \emph {et~al.}(1999)\citenamefont
  {Anishchenko}, \citenamefont {Neiman}, \citenamefont {Moss},\ and\
  \citenamefont {Shimansky-Geier}}]{Anishchenko1999}%
  \BibitemOpen
  \bibfield  {author} {\bibinfo {author} {\bibfnamefont {V.~S.}\ \bibnamefont
  {Anishchenko}}, \bibinfo {author} {\bibfnamefont {A.~B.}\ \bibnamefont
  {Neiman}}, \bibinfo {author} {\bibfnamefont {F.}~\bibnamefont {Moss}}, \ and\
  \bibinfo {author} {\bibfnamefont {L.}~\bibnamefont {Shimansky-Geier}},\
  }\href {\doibase 10.1070/pu1999v042n01abeh000444} {\bibfield  {journal}
  {\bibinfo  {journal} {Physics-Uspekhi}\ }\textbf {\bibinfo {volume} {42}},\
  \bibinfo {pages} {7} (\bibinfo {year} {1999})}\BibitemShut {NoStop}%
\bibitem [{\citenamefont {Wellens}\ \emph {et~al.}(2003)\citenamefont
  {Wellens}, \citenamefont {Shatokhin},\ and\ \citenamefont
  {Buchleitner}}]{Wellens2003}%
  \BibitemOpen
  \bibfield  {author} {\bibinfo {author} {\bibfnamefont {T.}~\bibnamefont
  {Wellens}}, \bibinfo {author} {\bibfnamefont {V.}~\bibnamefont {Shatokhin}},
  \ and\ \bibinfo {author} {\bibfnamefont {A.}~\bibnamefont {Buchleitner}},\
  }\href {\doibase 10.1088/0034-4885/67/1/r02} {\bibfield  {journal} {\bibinfo
  {journal} {Reports on Progress in Physics}\ }\textbf {\bibinfo {volume}
  {67}},\ \bibinfo {pages} {45} (\bibinfo {year} {2003})}\BibitemShut {NoStop}%
\bibitem [{\citenamefont {Benzi}(2010)}]{Benzi2010}%
  \BibitemOpen
  \bibfield  {author} {\bibinfo {author} {\bibfnamefont {R.}~\bibnamefont
  {Benzi}},\ }\href {\doibase 10.5194/npg-17-431-2010} {\bibfield  {journal}
  {\bibinfo  {journal} {Nonlinear Processes in Geophysics}\ }\textbf {\bibinfo
  {volume} {17}},\ \bibinfo {pages} {431} (\bibinfo {year} {2010})}\BibitemShut
  {NoStop}%
\bibitem [{\citenamefont {Freidlin}(2000)}]{Freidlin2000}%
  \BibitemOpen
  \bibfield  {author} {\bibinfo {author} {\bibfnamefont {M.~I.}\ \bibnamefont
  {Freidlin}},\ }\href {\doibase https://doi.org/10.1016/S0167-2789(99)00191-8}
  {\bibfield  {journal} {\bibinfo  {journal} {Physica D: Nonlinear Phenomena}\
  }\textbf {\bibinfo {volume} {137}},\ \bibinfo {pages} {333 } (\bibinfo {year}
  {2000})}\BibitemShut {NoStop}%
\bibitem [{\citenamefont {Herrmann}\ \emph {et~al.}(2005)\citenamefont
  {Herrmann}, \citenamefont {Imkeller},\ and\ \citenamefont
  {Pavlyukevich}}]{Herrmann2005}%
  \BibitemOpen
  \bibfield  {author} {\bibinfo {author} {\bibfnamefont {S.}~\bibnamefont
  {Herrmann}}, \bibinfo {author} {\bibfnamefont {P.}~\bibnamefont {Imkeller}},
  \ and\ \bibinfo {author} {\bibfnamefont {I.}~\bibnamefont {Pavlyukevich}},\
  }\enquote {\bibinfo {title} {Two mathematical approaches to stochastic
  resonance},}\ in\ \href@noop {} {\emph {\bibinfo {booktitle} {Interacting
  Stochastic Systems}}},\ \bibinfo {editor} {edited by\ \bibinfo {editor}
  {\bibfnamefont {J.~D.}\ \bibnamefont {Deuschel}}\ and\ \bibinfo {editor}
  {\bibfnamefont {A.}~\bibnamefont {Greven}}}\ (\bibinfo  {publisher}
  {Springer, Berlin},\ \bibinfo {year} {2005})\ pp.\ \bibinfo {pages}
  {327--351}\BibitemShut {NoStop}%
\bibitem [{\citenamefont {Imkeller}\ and\ \citenamefont
  {Pavlyukevich}(2002)}]{Imkeller2002}%
  \BibitemOpen
  \bibfield  {author} {\bibinfo {author} {\bibfnamefont {P.}~\bibnamefont
  {Imkeller}}\ and\ \bibinfo {author} {\bibfnamefont {I.}~\bibnamefont
  {Pavlyukevich}},\ }\href {\doibase 10.1142/S0219493702000583} {\bibfield
  {journal} {\bibinfo  {journal} {Stochastics and Dynamics}\ }\textbf {\bibinfo
  {volume} {02}},\ \bibinfo {pages} {463} (\bibinfo {year} {2002})},\ \Eprint
  {http://arxiv.org/abs/https://doi.org/10.1142/S0219493702000583}
  {https://doi.org/10.1142/S0219493702000583} \BibitemShut {NoStop}%
\bibitem [{\citenamefont {Imkeller}\ and\ \citenamefont
  {Pavlyukevich}(2004)}]{Imkeller2004}%
  \BibitemOpen
  \bibfield  {author} {\bibinfo {author} {\bibfnamefont {P.}~\bibnamefont
  {Imkeller}}\ and\ \bibinfo {author} {\bibfnamefont {I.}~\bibnamefont
  {Pavlyukevich}},\ }in\ \href@noop {} {\emph {\bibinfo {booktitle} {Seminar on
  Stochastic Analysis, Random Fields and Applications IV}}},\ \bibinfo {editor}
  {edited by\ \bibinfo {editor} {\bibfnamefont {R.~C.}\ \bibnamefont {Dalang}},
  \bibinfo {editor} {\bibfnamefont {M.}~\bibnamefont {Dozzi}}, \ and\ \bibinfo
  {editor} {\bibfnamefont {F.}~\bibnamefont {Russo}}}\ (\bibinfo  {publisher}
  {Birkh{\"a}user Basel},\ \bibinfo {address} {Basel},\ \bibinfo {year}
  {2004})\ pp.\ \bibinfo {pages} {141--154}\BibitemShut {NoStop}%
\bibitem [{\citenamefont {Herrmann}\ \emph {et~al.}(2014)\citenamefont
  {Herrmann}, \citenamefont {Imkeller}, \citenamefont {Pavlyukevich},\ and\
  \citenamefont {Peithmann}}]{Herrmann2014}%
  \BibitemOpen
  \bibfield  {author} {\bibinfo {author} {\bibfnamefont {S.}~\bibnamefont
  {Herrmann}}, \bibinfo {author} {\bibfnamefont {P.}~\bibnamefont {Imkeller}},
  \bibinfo {author} {\bibfnamefont {I.}~\bibnamefont {Pavlyukevich}}, \ and\
  \bibinfo {author} {\bibfnamefont {D.}~\bibnamefont {Peithmann}},\ }\href@noop
  {} {\emph {\bibinfo {title} {Stochastic Resonance: A Mathematical Approach in
  the Small Noise Limit}}},\ Mathematical Surveys and Monographs\ (\bibinfo
  {publisher} {AMS},\ \bibinfo {year} {2014})\BibitemShut {NoStop}%
\bibitem [{\citenamefont {Kramers}(1940)}]{Kramers1940}%
  \BibitemOpen
  \bibfield  {author} {\bibinfo {author} {\bibfnamefont {H.}~\bibnamefont
  {Kramers}},\ }\href {\doibase 10.1016/S0031-8914(40)90098-2} {\bibfield
  {journal} {\bibinfo  {journal} {Physica}\ }\textbf {\bibinfo {volume} {7}},\
  \bibinfo {pages} {284} (\bibinfo {year} {1940})}\BibitemShut {NoStop}%
\bibitem [{\citenamefont {Bovier}\ \emph {et~al.}(2004)\citenamefont {Bovier},
  \citenamefont {Eckhoff}, \citenamefont {Gayrard},\ and\ \citenamefont
  {Klein}}]{Bovier2004}%
  \BibitemOpen
  \bibfield  {author} {\bibinfo {author} {\bibfnamefont {A.}~\bibnamefont
  {Bovier}}, \bibinfo {author} {\bibfnamefont {M.}~\bibnamefont {Eckhoff}},
  \bibinfo {author} {\bibfnamefont {V.}~\bibnamefont {Gayrard}}, \ and\
  \bibinfo {author} {\bibfnamefont {M.}~\bibnamefont {Klein}},\ }\href
  {\doibase 10.4171/JEMS/14} {\bibfield  {journal} {\bibinfo  {journal}
  {Journal of the European Mathematical Society}\ }\textbf {\bibinfo {volume}
  {6}},\ \bibinfo {pages} {399} (\bibinfo {year} {2004})}\BibitemShut {NoStop}%
\bibitem [{\citenamefont {Berglund}(2013)}]{Berglund2013}%
  \BibitemOpen
  \bibfield  {author} {\bibinfo {author} {\bibfnamefont {N.}~\bibnamefont
  {Berglund}},\ }\href@noop {} {\bibfield  {journal} {\bibinfo  {journal}
  {Markov Process. Relat Fields}\ }\textbf {\bibinfo {volume} {19}},\ \bibinfo
  {pages} {459Ð490} (\bibinfo {year} {2013})}\BibitemShut {NoStop}%
\bibitem [{\citenamefont {Fox}\ and\ \citenamefont {Lu}(1993)}]{Fox1993}%
  \BibitemOpen
  \bibfield  {author} {\bibinfo {author} {\bibfnamefont {R.~F.}\ \bibnamefont
  {Fox}}\ and\ \bibinfo {author} {\bibfnamefont {Y.-n.}\ \bibnamefont {Lu}},\
  }\href {\doibase 10.1103/PhysRevE.48.3390} {\bibfield  {journal} {\bibinfo
  {journal} {Phys. Rev. E}\ }\textbf {\bibinfo {volume} {48}},\ \bibinfo
  {pages} {3390} (\bibinfo {year} {1993})}\BibitemShut {NoStop}%
\bibitem [{\citenamefont {Tessone}\ and\ \citenamefont
  {Wio}(1998)}]{Tessone1998}%
  \BibitemOpen
  \bibfield  {author} {\bibinfo {author} {\bibfnamefont {C.~J.}\ \bibnamefont
  {Tessone}}\ and\ \bibinfo {author} {\bibfnamefont {H.~S.}\ \bibnamefont
  {Wio}},\ }\href {\doibase 10.1142/S0217984998001414} {\bibfield  {journal}
  {\bibinfo  {journal} {Modern Physics Letters B}\ }\textbf {\bibinfo {volume}
  {12}},\ \bibinfo {pages} {1195} (\bibinfo {year} {1998})},\ \Eprint
  {http://arxiv.org/abs/https://doi.org/10.1142/S0217984998001414}
  {https://doi.org/10.1142/S0217984998001414} \BibitemShut {NoStop}%
\bibitem [{\citenamefont {Jia}\ \emph {et~al.}(2000)\citenamefont {Jia},
  \citenamefont {Yu},\ and\ \citenamefont {Li}}]{Jia2000}%
  \BibitemOpen
  \bibfield  {author} {\bibinfo {author} {\bibfnamefont {Y.}~\bibnamefont
  {Jia}}, \bibinfo {author} {\bibfnamefont {S.-n.}\ \bibnamefont {Yu}}, \ and\
  \bibinfo {author} {\bibfnamefont {J.-r.}\ \bibnamefont {Li}},\ }\href
  {\doibase 10.1103/PhysRevE.62.1869} {\bibfield  {journal} {\bibinfo
  {journal} {Phys. Rev. E}\ }\textbf {\bibinfo {volume} {62}},\ \bibinfo
  {pages} {1869} (\bibinfo {year} {2000})}\BibitemShut {NoStop}%
\bibitem [{\citenamefont {Jia}\ \emph {et~al.}(2001)\citenamefont {Jia},
  \citenamefont {Zheng}, \citenamefont {Hu},\ and\ \citenamefont
  {Li}}]{Jia2001}%
  \BibitemOpen
  \bibfield  {author} {\bibinfo {author} {\bibfnamefont {Y.}~\bibnamefont
  {Jia}}, \bibinfo {author} {\bibfnamefont {X.-p.}\ \bibnamefont {Zheng}},
  \bibinfo {author} {\bibfnamefont {X.-m.}\ \bibnamefont {Hu}}, \ and\ \bibinfo
  {author} {\bibfnamefont {J.-r.}\ \bibnamefont {Li}},\ }\href {\doibase
  10.1103/PhysRevE.63.031107} {\bibfield  {journal} {\bibinfo  {journal} {Phys.
  Rev. E}\ }\textbf {\bibinfo {volume} {63}},\ \bibinfo {pages} {031107}
  (\bibinfo {year} {2001})}\BibitemShut {NoStop}%
\bibitem [{\citenamefont {Kuhwald}\ and\ \citenamefont
  {Pavlyukevich}(2016)}]{Kuhwald2016}%
  \BibitemOpen
  \bibfield  {author} {\bibinfo {author} {\bibfnamefont {I.}~\bibnamefont
  {Kuhwald}}\ and\ \bibinfo {author} {\bibfnamefont {I.}~\bibnamefont
  {Pavlyukevich}},\ }\href {\doibase
  https://doi.org/10.1016/j.proeng.2016.05.129} {\bibfield  {journal} {\bibinfo
   {journal} {Procedia Engineering}\ }\textbf {\bibinfo {volume} {144}},\
  \bibinfo {pages} {1307 } (\bibinfo {year} {2016})},\ \bibinfo {note}
  {international Conference on Vibration Problems 2015}\BibitemShut {NoStop}%
\bibitem [{\citenamefont {Qiao}\ \emph {et~al.}(2016)\citenamefont {Qiao},
  \citenamefont {Lei}, \citenamefont {Lin},\ and\ \citenamefont
  {Niu}}]{Qiao2016}%
  \BibitemOpen
  \bibfield  {author} {\bibinfo {author} {\bibfnamefont {Z.}~\bibnamefont
  {Qiao}}, \bibinfo {author} {\bibfnamefont {Y.}~\bibnamefont {Lei}}, \bibinfo
  {author} {\bibfnamefont {J.}~\bibnamefont {Lin}}, \ and\ \bibinfo {author}
  {\bibfnamefont {S.}~\bibnamefont {Niu}},\ }\href {\doibase
  10.1103/PhysRevE.94.052214} {\bibfield  {journal} {\bibinfo  {journal} {Phys.
  Rev. E}\ }\textbf {\bibinfo {volume} {94}},\ \bibinfo {pages} {052214}
  (\bibinfo {year} {2016})}\BibitemShut {NoStop}%
\bibitem [{\citenamefont {McNamara}\ and\ \citenamefont
  {Wiesenfeld}(1989)}]{McNamara1989}%
  \BibitemOpen
  \bibfield  {author} {\bibinfo {author} {\bibfnamefont {B.}~\bibnamefont
  {McNamara}}\ and\ \bibinfo {author} {\bibfnamefont {K.}~\bibnamefont
  {Wiesenfeld}},\ }\href {\doibase 10.1103/PhysRevA.39.4854} {\bibfield
  {journal} {\bibinfo  {journal} {Phys. Rev. A}\ }\textbf {\bibinfo {volume}
  {39}},\ \bibinfo {pages} {4854} (\bibinfo {year} {1989})}\BibitemShut
  {NoStop}%
\bibitem [{\citenamefont {Leli{\`e}vre}(2015)}]{Lelievre2015}%
  \BibitemOpen
  \bibfield  {author} {\bibinfo {author} {\bibfnamefont {T.}~\bibnamefont
  {Leli{\`e}vre}},\ }\href {\doibase 10.1140/epjst/e2015-02420-1} {\bibfield
  {journal} {\bibinfo  {journal} {The European Physical Journal Special
  Topics}\ }\textbf {\bibinfo {volume} {224}},\ \bibinfo {pages} {2429}
  (\bibinfo {year} {2015})}\BibitemShut {NoStop}%
\bibitem [{\citenamefont {Ges{\`u}}\ \emph {et~al.}(2019)\citenamefont
  {Ges{\`u}}, \citenamefont {Leli{\`e}vre}, \citenamefont {Peutrec},\ and\
  \citenamefont {Nectoux}}]{Gesu2019}%
  \BibitemOpen
  \bibfield  {author} {\bibinfo {author} {\bibfnamefont {G.~D.}\ \bibnamefont
  {Ges{\`u}}}, \bibinfo {author} {\bibfnamefont {T.}~\bibnamefont
  {Leli{\`e}vre}}, \bibinfo {author} {\bibfnamefont {D.~L.}\ \bibnamefont
  {Peutrec}}, \ and\ \bibinfo {author} {\bibfnamefont {B.}~\bibnamefont
  {Nectoux}},\ }\href {\doibase 10.1007/s40818-019-0059-2} {\bibfield
  {journal} {\bibinfo  {journal} {Annals of PDE}\ }\textbf {\bibinfo {volume}
  {5}},\ \bibinfo {pages} {5} (\bibinfo {year} {2019})}\BibitemShut {NoStop}%
\bibitem [{\citenamefont {Bouzat}\ and\ \citenamefont
  {Wio}(1999)}]{Bouzat1999}%
  \BibitemOpen
  \bibfield  {author} {\bibinfo {author} {\bibfnamefont {S.}~\bibnamefont
  {Bouzat}}\ and\ \bibinfo {author} {\bibfnamefont {H.~S.}\ \bibnamefont
  {Wio}},\ }\href {\doibase 10.1103/PhysRevE.59.5142} {\bibfield  {journal}
  {\bibinfo  {journal} {Phys. Rev. E}\ }\textbf {\bibinfo {volume} {59}},\
  \bibinfo {pages} {5142} (\bibinfo {year} {1999})}\BibitemShut {NoStop}%
\bibitem [{\citenamefont {Wio}\ and\ \citenamefont {Bouzat}(1999)}]{Wio1999}%
  \BibitemOpen
  \bibfield  {author} {\bibinfo {author} {\bibfnamefont {H.~S.}\ \bibnamefont
  {Wio}}\ and\ \bibinfo {author} {\bibfnamefont {S.~A.}\ \bibnamefont
  {Bouzat}},\ }\href@noop {} {\bibfield  {journal} {\bibinfo  {journal}
  {{Brazilian Journal of Physics}}\ }\textbf {\bibinfo {volume} {29}},\
  \bibinfo {pages} {136 } (\bibinfo {year} {1999})}\BibitemShut {NoStop}%
\bibitem [{\citenamefont {Grebogi}\ \emph {et~al.}(1983)\citenamefont
  {Grebogi}, \citenamefont {Ott},\ and\ \citenamefont {Yorke}}]{Grebogi1983}%
  \BibitemOpen
  \bibfield  {author} {\bibinfo {author} {\bibfnamefont {C.}~\bibnamefont
  {Grebogi}}, \bibinfo {author} {\bibfnamefont {E.}~\bibnamefont {Ott}}, \ and\
  \bibinfo {author} {\bibfnamefont {J.~A.}\ \bibnamefont {Yorke}},\ }\href
  {\doibase 10.1103/PhysRevLett.50.935} {\bibfield  {journal} {\bibinfo
  {journal} {Physical Review Letters}\ }\textbf {\bibinfo {volume} {50}},\
  \bibinfo {pages} {935} (\bibinfo {year} {1983})}\BibitemShut {NoStop}%
\bibitem [{\citenamefont {Robert}\ \emph {et~al.}(2000)\citenamefont {Robert},
  \citenamefont {Alligood}, \citenamefont {Ott},\ and\ \citenamefont
  {Yorke}}]{Robert2000}%
  \BibitemOpen
  \bibfield  {author} {\bibinfo {author} {\bibfnamefont {C.}~\bibnamefont
  {Robert}}, \bibinfo {author} {\bibfnamefont {K.~T.}\ \bibnamefont
  {Alligood}}, \bibinfo {author} {\bibfnamefont {E.}~\bibnamefont {Ott}}, \
  and\ \bibinfo {author} {\bibfnamefont {J.~A.}\ \bibnamefont {Yorke}},\ }\href
  {\doibase 10.1016/S0167-2789(00)00074-9} {\bibfield  {journal} {\bibinfo
  {journal} {Physica D: Nonlinear Phenomena}\ }\textbf {\bibinfo {volume}
  {144}},\ \bibinfo {pages} {44} (\bibinfo {year} {2000})}\BibitemShut
  {NoStop}%
\bibitem [{\citenamefont {Ott}(2002)}]{Ott2002}%
  \BibitemOpen
  \bibfield  {author} {\bibinfo {author} {\bibfnamefont {E.}~\bibnamefont
  {Ott}},\ }\href@noop {} {\emph {\bibinfo {title} {Chaos in Dynamical
  Systems}}}\ (\bibinfo  {publisher} {Cambridge University Press},\ \bibinfo
  {year} {2002})\BibitemShut {NoStop}%
\bibitem [{\citenamefont {Vollmer}\ \emph {et~al.}(2009)\citenamefont
  {Vollmer}, \citenamefont {Schneider},\ and\ \citenamefont
  {Eckhardt}}]{Vollmer2009}%
  \BibitemOpen
  \bibfield  {author} {\bibinfo {author} {\bibfnamefont {J.}~\bibnamefont
  {Vollmer}}, \bibinfo {author} {\bibfnamefont {T.~M.}\ \bibnamefont
  {Schneider}}, \ and\ \bibinfo {author} {\bibfnamefont {B.}~\bibnamefont
  {Eckhardt}},\ }\href@noop {} {\bibfield  {journal} {\bibinfo  {journal} {New
  Journal of Physics}\ }\textbf {\bibinfo {volume} {11}},\ \bibinfo {pages}
  {013040} (\bibinfo {year} {2009})}\BibitemShut {NoStop}%
\bibitem [{\citenamefont {Lucarini}\ and\ \citenamefont
  {B{\'{o}}dai}(2017)}]{LucariniBodai2017}%
  \BibitemOpen
  \bibfield  {author} {\bibinfo {author} {\bibfnamefont {V.}~\bibnamefont
  {Lucarini}}\ and\ \bibinfo {author} {\bibfnamefont {T.}~\bibnamefont
  {B{\'{o}}dai}},\ }\href {\doibase 10.1088/1361-6544/aa6b11} {\bibfield
  {journal} {\bibinfo  {journal} {Nonlinearity}\ }\textbf {\bibinfo {volume}
  {30}},\ \bibinfo {pages} {R32} (\bibinfo {year} {2017})}\BibitemShut
  {NoStop}%
\bibitem [{\citenamefont {Nicolis}\ \emph {et~al.}(1993)\citenamefont
  {Nicolis}, \citenamefont {Nicolis},\ and\ \citenamefont
  {McKernan}}]{Nicolis1993}%
  \BibitemOpen
  \bibfield  {author} {\bibinfo {author} {\bibfnamefont {G.}~\bibnamefont
  {Nicolis}}, \bibinfo {author} {\bibfnamefont {C.}~\bibnamefont {Nicolis}}, \
  and\ \bibinfo {author} {\bibfnamefont {D.}~\bibnamefont {McKernan}},\ }\href
  {\doibase 10.1007/BF01053958} {\bibfield  {journal} {\bibinfo  {journal}
  {Journal of Statistical Physics}\ }\textbf {\bibinfo {volume} {70}},\
  \bibinfo {pages} {125} (\bibinfo {year} {1993})}\BibitemShut {NoStop}%
\bibitem [{\citenamefont {Anishchenko}\ \emph {et~al.}(1993)\citenamefont
  {Anishchenko}, \citenamefont {Neiman},\ and\ \citenamefont
  {Safanova}}]{Anishchenko1993}%
  \BibitemOpen
  \bibfield  {author} {\bibinfo {author} {\bibfnamefont {V.~S.}\ \bibnamefont
  {Anishchenko}}, \bibinfo {author} {\bibfnamefont {A.~B.}\ \bibnamefont
  {Neiman}}, \ and\ \bibinfo {author} {\bibfnamefont {M.~A.}\ \bibnamefont
  {Safanova}},\ }\href {\doibase 10.1007/BF01053962} {\bibfield  {journal}
  {\bibinfo  {journal} {Journal of Statistical Physics}\ }\textbf {\bibinfo
  {volume} {70}},\ \bibinfo {pages} {183} (\bibinfo {year} {1993})}\BibitemShut
  {NoStop}%
\bibitem [{\citenamefont {Crisanti}\ \emph {et~al.}(1994)\citenamefont
  {Crisanti}, \citenamefont {Falcioni}, \citenamefont {Paladin},\ and\
  \citenamefont {Vulpiani}}]{Crisanti1994}%
  \BibitemOpen
  \bibfield  {author} {\bibinfo {author} {\bibfnamefont {A.}~\bibnamefont
  {Crisanti}}, \bibinfo {author} {\bibfnamefont {M.}~\bibnamefont {Falcioni}},
  \bibinfo {author} {\bibfnamefont {G.}~\bibnamefont {Paladin}}, \ and\
  \bibinfo {author} {\bibfnamefont {A.}~\bibnamefont {Vulpiani}},\ }\href
  {\doibase 10.1088/0305-4470/27/17/001} {\bibfield  {journal} {\bibinfo
  {journal} {Journal of Physics A: Mathematical and General}\ }\textbf
  {\bibinfo {volume} {27}},\ \bibinfo {pages} {L597} (\bibinfo {year}
  {1994})}\BibitemShut {NoStop}%
\bibitem [{\citenamefont {Lucarini}\ and\ \citenamefont
  {B\'odai}(019a)}]{Lucarini2019}%
  \BibitemOpen
  \bibfield  {author} {\bibinfo {author} {\bibfnamefont {V.}~\bibnamefont
  {Lucarini}}\ and\ \bibinfo {author} {\bibfnamefont {T.}~\bibnamefont
  {B\'odai}},\ }\href {\doibase 10.1103/PhysRevLett.122.158701} {\bibfield
  {journal} {\bibinfo  {journal} {Phys. Rev. Lett.}\ }\textbf {\bibinfo
  {volume} {122}},\ \bibinfo {pages} {158701} (\bibinfo {year}
  {2019a})}\BibitemShut {NoStop}%
\bibitem [{\citenamefont {Lucarini}\ and\ \citenamefont
  {B\'odai}(019b)}]{LucariniBodai2019arxiv}%
  \BibitemOpen
  \bibfield  {author} {\bibinfo {author} {\bibfnamefont {V.}~\bibnamefont
  {Lucarini}}\ and\ \bibinfo {author} {\bibfnamefont {T.}~\bibnamefont
  {B\'odai}},\ }\href@noop {} {\bibfield  {journal} {\bibinfo  {journal} {arXiv
  e-prints}\ } (\bibinfo {year} {2019b})},\ \Eprint
  {http://arxiv.org/abs/1903.08348} {arXiv:1903.08348 [physics.ao-ph]}
  \BibitemShut {NoStop}%
\bibitem [{\citenamefont {Freidlin}\ and\ \citenamefont
  {Wentzell}(1984)}]{Freidlin1984}%
  \BibitemOpen
  \bibfield  {author} {\bibinfo {author} {\bibfnamefont {M.~I.}\ \bibnamefont
  {Freidlin}}\ and\ \bibinfo {author} {\bibfnamefont {A.}~\bibnamefont
  {Wentzell}},\ }\href@noop {} {\emph {\bibinfo {title} {Random Perturbations
  of Dynamical Systems}}}\ (\bibinfo  {publisher} {Springer},\ \bibinfo
  {address} {New York},\ \bibinfo {year} {1984})\BibitemShut {NoStop}%
\bibitem [{\citenamefont {Graham}\ \emph {et~al.}(1991)\citenamefont {Graham},
  \citenamefont {Hamm},\ and\ \citenamefont {T\'el}}]{Graham1991}%
  \BibitemOpen
  \bibfield  {author} {\bibinfo {author} {\bibfnamefont {R.}~\bibnamefont
  {Graham}}, \bibinfo {author} {\bibfnamefont {A.}~\bibnamefont {Hamm}}, \ and\
  \bibinfo {author} {\bibfnamefont {T.}~\bibnamefont {T\'el}},\ }\href
  {\doibase 10.1103/PhysRevLett.66.3089} {\bibfield  {journal} {\bibinfo
  {journal} {Phys. Rev. Lett.}\ }\textbf {\bibinfo {volume} {66}},\ \bibinfo
  {pages} {3089} (\bibinfo {year} {1991})}\BibitemShut {NoStop}%
\bibitem [{\citenamefont {Hamm}\ \emph {et~al.}(1994)\citenamefont {Hamm},
  \citenamefont {T\'el},\ and\ \citenamefont {Graham}}]{Hamm1994}%
  \BibitemOpen
  \bibfield  {author} {\bibinfo {author} {\bibfnamefont {A.}~\bibnamefont
  {Hamm}}, \bibinfo {author} {\bibfnamefont {T.}~\bibnamefont {T\'el}}, \ and\
  \bibinfo {author} {\bibfnamefont {R.}~\bibnamefont {Graham}},\ }\href
  {\doibase 10.1016/0375-9601(94)90621-1} {\bibfield  {journal} {\bibinfo
  {journal} {Physics Letters A}\ }\textbf {\bibinfo {volume} {185}},\ \bibinfo
  {pages} {313} (\bibinfo {year} {1994})}\BibitemShut {NoStop}%
\bibitem [{\citenamefont {Lai}\ and\ \citenamefont {T\'el}(2011)}]{LT:2011}%
  \BibitemOpen
  \bibfield  {author} {\bibinfo {author} {\bibfnamefont {Y.-C.}\ \bibnamefont
  {Lai}}\ and\ \bibinfo {author} {\bibfnamefont {T.}~\bibnamefont {T\'el}},\
  }\href@noop {} {\emph {\bibinfo {title} {Transient Chaos}}}\ (\bibinfo
  {publisher} {Springer},\ \bibinfo {address} {New York},\ \bibinfo {year}
  {2011})\BibitemShut {NoStop}%
\bibitem [{\citenamefont {Gaspard}(2002)}]{Gaspard2002}%
  \BibitemOpen
  \bibfield  {author} {\bibinfo {author} {\bibfnamefont {P.}~\bibnamefont
  {Gaspard}},\ }\href {\doibase 10.1023/A:1013167928166} {\bibfield  {journal}
  {\bibinfo  {journal} {Journal of Statistical Physics}\ }\textbf {\bibinfo
  {volume} {106}},\ \bibinfo {pages} {57} (\bibinfo {year} {2002})}\BibitemShut
  {NoStop}%
\bibitem [{\citenamefont {Bouchet}\ \emph {et~al.}(2016)\citenamefont
  {Bouchet}, \citenamefont {Gawedzki},\ and\ \citenamefont
  {Nardini}}]{Nardini2016}%
  \BibitemOpen
  \bibfield  {author} {\bibinfo {author} {\bibfnamefont {F.}~\bibnamefont
  {Bouchet}}, \bibinfo {author} {\bibfnamefont {K.}~\bibnamefont {Gawedzki}}, \
  and\ \bibinfo {author} {\bibfnamefont {C.}~\bibnamefont {Nardini}},\ }\href
  {\doibase 10.1007/s10955-016-1503-2} {\bibfield  {journal} {\bibinfo
  {journal} {Journal of Statistical Physics}\ }\textbf {\bibinfo {volume}
  {163}},\ \bibinfo {pages} {1157} (\bibinfo {year} {2016})}\BibitemShut
  {NoStop}%
\bibitem [{\citenamefont {Ao}(2004)}]{Ao2004}%
  \BibitemOpen
  \bibfield  {author} {\bibinfo {author} {\bibfnamefont {P.}~\bibnamefont
  {Ao}},\ }\href {\doibase 10.1088/0305-4470/37/3/l01} {\bibfield  {journal}
  {\bibinfo  {journal} {Journal of Physics A: Mathematical and General}\
  }\textbf {\bibinfo {volume} {37}},\ \bibinfo {pages} {L25} (\bibinfo {year}
  {2004})}\BibitemShut {NoStop}%
\bibitem [{\citenamefont {Yin}\ and\ \citenamefont {Ao}(2006)}]{Yin2006}%
  \BibitemOpen
  \bibfield  {author} {\bibinfo {author} {\bibfnamefont {L.}~\bibnamefont
  {Yin}}\ and\ \bibinfo {author} {\bibfnamefont {P.}~\bibnamefont {Ao}},\
  }\href {\doibase 10.1088/0305-4470/39/27/003} {\bibfield  {journal} {\bibinfo
   {journal} {Journal of Physics A: Mathematical and General}\ }\textbf
  {\bibinfo {volume} {39}},\ \bibinfo {pages} {8593} (\bibinfo {year}
  {2006})}\BibitemShut {NoStop}%
\bibitem [{\citenamefont {Graham}\ and\ \citenamefont
  {T\'el}(1986)}]{Graham1986}%
  \BibitemOpen
  \bibfield  {author} {\bibinfo {author} {\bibfnamefont {R.}~\bibnamefont
  {Graham}}\ and\ \bibinfo {author} {\bibfnamefont {T.}~\bibnamefont {T\'el}},\
  }\href {\doibase 10.1103/PhysRevA.33.1322} {\bibfield  {journal} {\bibinfo
  {journal} {Phys. Rev. A}\ }\textbf {\bibinfo {volume} {33}},\ \bibinfo
  {pages} {1322} (\bibinfo {year} {1986})}\BibitemShut {NoStop}%
\bibitem [{\citenamefont {Zhou}\ \emph {et~al.}(2012)\citenamefont {Zhou},
  \citenamefont {Aliyu}, \citenamefont {Aurell},\ and\ \citenamefont
  {Huang}}]{Zhou2012}%
  \BibitemOpen
  \bibfield  {author} {\bibinfo {author} {\bibfnamefont {J.~X.}\ \bibnamefont
  {Zhou}}, \bibinfo {author} {\bibfnamefont {M.~D.~S.}\ \bibnamefont {Aliyu}},
  \bibinfo {author} {\bibfnamefont {E.}~\bibnamefont {Aurell}}, \ and\ \bibinfo
  {author} {\bibfnamefont {S.}~\bibnamefont {Huang}},\ }\href {\doibase
  10.1098/rsif.2012.0434} {\bibfield  {journal} {\bibinfo  {journal} {Journal
  of The Royal Society Interface}\ }\textbf {\bibinfo {volume} {9}},\ \bibinfo
  {pages} {3539} (\bibinfo {year} {2012})},\ \Eprint
  {http://arxiv.org/abs/https://royalsocietypublishing.org/doi/pdf/10.1098/rsif.2012.0434}
  {https://royalsocietypublishing.org/doi/pdf/10.1098/rsif.2012.0434}
  \BibitemShut {NoStop}%
\bibitem [{\citenamefont {Brackston}\ \emph {et~al.}(2018)\citenamefont
  {Brackston}, \citenamefont {Wynn},\ and\ \citenamefont
  {Stumpf}}]{Brackston2018}%
  \BibitemOpen
  \bibfield  {author} {\bibinfo {author} {\bibfnamefont {R.~D.}\ \bibnamefont
  {Brackston}}, \bibinfo {author} {\bibfnamefont {A.}~\bibnamefont {Wynn}}, \
  and\ \bibinfo {author} {\bibfnamefont {M.~P.~H.}\ \bibnamefont {Stumpf}},\
  }\href {\doibase 10.1103/PhysRevE.98.022136} {\bibfield  {journal} {\bibinfo
  {journal} {Phys. Rev. E}\ }\textbf {\bibinfo {volume} {98}},\ \bibinfo
  {pages} {022136} (\bibinfo {year} {2018})}\BibitemShut {NoStop}%
\bibitem [{\citenamefont {Tang}\ \emph {et~al.}(2017)\citenamefont {Tang},
  \citenamefont {Yuan}, \citenamefont {Wang}, \citenamefont {Zhu},\ and\
  \citenamefont {Ao}}]{Tang2017}%
  \BibitemOpen
  \bibfield  {author} {\bibinfo {author} {\bibfnamefont {Y.}~\bibnamefont
  {Tang}}, \bibinfo {author} {\bibfnamefont {R.}~\bibnamefont {Yuan}}, \bibinfo
  {author} {\bibfnamefont {G.}~\bibnamefont {Wang}}, \bibinfo {author}
  {\bibfnamefont {X.}~\bibnamefont {Zhu}}, \ and\ \bibinfo {author}
  {\bibfnamefont {P.}~\bibnamefont {Ao}},\ }\href {\doibase
  10.1038/s41598-017-15889-2} {\bibfield  {journal} {\bibinfo  {journal}
  {Scientific Reports}\ }\textbf {\bibinfo {volume} {7}},\ \bibinfo {pages}
  {15762} (\bibinfo {year} {2017})}\BibitemShut {NoStop}%
\bibitem [{\citenamefont {Graham}(1987)}]{Graham1987}%
  \BibitemOpen
  \bibfield  {author} {\bibinfo {author} {\bibfnamefont {R.}~\bibnamefont
  {Graham}},\ }in\ \href@noop {} {\emph {\bibinfo {booktitle} {Fluctuations and
  Stochastic Phenomena in Condensed Matter}}},\ \bibinfo {editor} {edited by\
  \bibinfo {editor} {\bibfnamefont {L.}~\bibnamefont {Garrido}}}\ (\bibinfo
  {publisher} {Springer Berlin Heidelberg},\ \bibinfo {address} {Berlin,
  Heidelberg},\ \bibinfo {year} {1987})\ pp.\ \bibinfo {pages}
  {1--34}\BibitemShut {NoStop}%
\bibitem [{\citenamefont {Bouchet}\ and\ \citenamefont
  {Reygner}(2016)}]{Bouchet2016}%
  \BibitemOpen
  \bibfield  {author} {\bibinfo {author} {\bibfnamefont {F.}~\bibnamefont
  {Bouchet}}\ and\ \bibinfo {author} {\bibfnamefont {J.}~\bibnamefont
  {Reygner}},\ }\href {\doibase 10.1007/s00023-016-0507-4} {\bibfield
  {journal} {\bibinfo  {journal} {Annales Henri Poincar{\'e}}\ }\textbf
  {\bibinfo {volume} {17}},\ \bibinfo {pages} {3499} (\bibinfo {year}
  {2016})}\BibitemShut {NoStop}%
\bibitem [{\citenamefont {T{\u{a}}nase-Nicola}\ and\ \citenamefont
  {Kurchan}(2004)}]{Tanase-Nicola2004}%
  \BibitemOpen
  \bibfield  {author} {\bibinfo {author} {\bibfnamefont {S.}~\bibnamefont
  {T{\u{a}}nase-Nicola}}\ and\ \bibinfo {author} {\bibfnamefont
  {J.}~\bibnamefont {Kurchan}},\ }\href {\doibase
  10.1023/B:JOSS.0000041739.53068.6a} {\bibfield  {journal} {\bibinfo
  {journal} {Journal of Statistical Physics}\ }\textbf {\bibinfo {volume}
  {116}},\ \bibinfo {pages} {1201} (\bibinfo {year} {2004})}\BibitemShut
  {NoStop}%
\bibitem [{\citenamefont {Lucarini}\ \emph {et~al.}(2012)\citenamefont
  {Lucarini}, \citenamefont {Faranda},\ and\ \citenamefont
  {Willeit}}]{LucariniFarandaWilleit2012}%
  \BibitemOpen
  \bibfield  {author} {\bibinfo {author} {\bibfnamefont {V.}~\bibnamefont
  {Lucarini}}, \bibinfo {author} {\bibfnamefont {D.}~\bibnamefont {Faranda}}, \
  and\ \bibinfo {author} {\bibfnamefont {M.}~\bibnamefont {Willeit}},\ }\href
  {\doibase 10.5194/npg-19-9-2012} {\bibfield  {journal} {\bibinfo  {journal}
  {Nonlinear Processes in Geophysics}\ }\textbf {\bibinfo {volume} {19}},\
  \bibinfo {pages} {9} (\bibinfo {year} {2012})}\BibitemShut {NoStop}%
\bibitem [{\citenamefont {Lenton}\ \emph {et~al.}(2008)\citenamefont {Lenton},
  \citenamefont {Held}, \citenamefont {Kriegler}, \citenamefont {Hall},
  \citenamefont {Lucht}, \citenamefont {Rahmstorf},\ and\ \citenamefont
  {Schellnhuber}}]{Lenton2008}%
  \BibitemOpen
  \bibfield  {author} {\bibinfo {author} {\bibfnamefont {T.~M.}\ \bibnamefont
  {Lenton}}, \bibinfo {author} {\bibfnamefont {H.}~\bibnamefont {Held}},
  \bibinfo {author} {\bibfnamefont {E.}~\bibnamefont {Kriegler}}, \bibinfo
  {author} {\bibfnamefont {J.~W.}\ \bibnamefont {Hall}}, \bibinfo {author}
  {\bibfnamefont {W.}~\bibnamefont {Lucht}}, \bibinfo {author} {\bibfnamefont
  {S.}~\bibnamefont {Rahmstorf}}, \ and\ \bibinfo {author} {\bibfnamefont
  {H.~J.}\ \bibnamefont {Schellnhuber}},\ }\href {\doibase
  10.1073/pnas.0705414105} {\bibfield  {journal} {\bibinfo  {journal}
  {Proceedings of the National Academy of Sciences}\ }\textbf {\bibinfo
  {volume} {105}},\ \bibinfo {pages} {1786} (\bibinfo {year}
  {2008})}\BibitemShut {NoStop}%
\bibitem [{\citenamefont {Ghil}\ and\ \citenamefont
  {Lucarini}(2019)}]{Ghil2019}%
  \BibitemOpen
  \bibfield  {author} {\bibinfo {author} {\bibfnamefont {M.}~\bibnamefont
  {Ghil}}\ and\ \bibinfo {author} {\bibfnamefont {V.}~\bibnamefont
  {Lucarini}},\ }\href@noop {} {\enquote {\bibinfo {title} {The physics of
  climate variability and climate change},}\ } (\bibinfo {year} {2019}),\
  \Eprint {http://arxiv.org/abs/1910.00583} {arXiv:1910.00583 [physics.ao-ph]}
  \BibitemShut {NoStop}%
\bibitem [{\citenamefont {Rahmstorf}\ \emph {et~al.}(2005)\citenamefont
  {Rahmstorf}, \citenamefont {Crucifix}, \citenamefont {Ganopolski},
  \citenamefont {Goosse}, \citenamefont {Kamenkovich}, \citenamefont {Knutti},
  \citenamefont {Lohmann}, \citenamefont {Marsh}, \citenamefont {Mysak},
  \citenamefont {Wang},\ and\ \citenamefont {Weaver}}]{Rahmstorf2005}%
  \BibitemOpen
  \bibfield  {author} {\bibinfo {author} {\bibfnamefont {S.}~\bibnamefont
  {Rahmstorf}}, \bibinfo {author} {\bibfnamefont {M.}~\bibnamefont {Crucifix}},
  \bibinfo {author} {\bibfnamefont {A.}~\bibnamefont {Ganopolski}}, \bibinfo
  {author} {\bibfnamefont {H.}~\bibnamefont {Goosse}}, \bibinfo {author}
  {\bibfnamefont {I.}~\bibnamefont {Kamenkovich}}, \bibinfo {author}
  {\bibfnamefont {R.}~\bibnamefont {Knutti}}, \bibinfo {author} {\bibfnamefont
  {G.}~\bibnamefont {Lohmann}}, \bibinfo {author} {\bibfnamefont
  {R.}~\bibnamefont {Marsh}}, \bibinfo {author} {\bibfnamefont {L.~A.}\
  \bibnamefont {Mysak}}, \bibinfo {author} {\bibfnamefont {Z.}~\bibnamefont
  {Wang}}, \ and\ \bibinfo {author} {\bibfnamefont {A.~J.}\ \bibnamefont
  {Weaver}},\ }\href {\doibase 10.1029/2005GL023655} {\bibfield  {journal}
  {\bibinfo  {journal} {Geophysical Research Letters}\ }\textbf {\bibinfo
  {volume} {32}} (\bibinfo {year} {2005}),\ 10.1029/2005GL023655}\BibitemShut
  {NoStop}%
\bibitem [{\citenamefont {Kuhlbrodt}\ \emph {et~al.}(2007)\citenamefont
  {Kuhlbrodt}, \citenamefont {Griesel}, \citenamefont {Montoya}, \citenamefont
  {Levermann}, \citenamefont {Hofmann},\ and\ \citenamefont
  {Rahmstorf}}]{Kuhlbrodt2007}%
  \BibitemOpen
  \bibfield  {author} {\bibinfo {author} {\bibfnamefont {T.}~\bibnamefont
  {Kuhlbrodt}}, \bibinfo {author} {\bibfnamefont {A.}~\bibnamefont {Griesel}},
  \bibinfo {author} {\bibfnamefont {M.}~\bibnamefont {Montoya}}, \bibinfo
  {author} {\bibfnamefont {A.}~\bibnamefont {Levermann}}, \bibinfo {author}
  {\bibfnamefont {M.}~\bibnamefont {Hofmann}}, \ and\ \bibinfo {author}
  {\bibfnamefont {S.}~\bibnamefont {Rahmstorf}},\ }\href {\doibase
  10.1029/2004RG000166} {\bibfield  {journal} {\bibinfo  {journal} {Reviews of
  Geophysics}\ }\textbf {\bibinfo {volume} {45}} (\bibinfo {year} {2007}),\
  10.1029/2004RG000166},\ \Eprint
  {http://arxiv.org/abs/https://agupubs.onlinelibrary.wiley.com/doi/pdf/10.1029/2004RG000166}
  {https://agupubs.onlinelibrary.wiley.com/doi/pdf/10.1029/2004RG000166}
  \BibitemShut {NoStop}%
\bibitem [{\citenamefont {Choi}\ \emph {et~al.}(1998)\citenamefont {Choi},
  \citenamefont {Fox},\ and\ \citenamefont {Jung}}]{Choi1998}%
  \BibitemOpen
  \bibfield  {author} {\bibinfo {author} {\bibfnamefont {M.~H.}\ \bibnamefont
  {Choi}}, \bibinfo {author} {\bibfnamefont {R.~F.}\ \bibnamefont {Fox}}, \
  and\ \bibinfo {author} {\bibfnamefont {P.}~\bibnamefont {Jung}},\ }\href
  {\doibase 10.1103/PhysRevE.57.6335} {\bibfield  {journal} {\bibinfo
  {journal} {Phys. Rev. E}\ }\textbf {\bibinfo {volume} {57}},\ \bibinfo
  {pages} {6335} (\bibinfo {year} {1998})}\BibitemShut {NoStop}%
\bibitem [{\citenamefont {Berglund}\ and\ \citenamefont
  {Gentz}(2005)}]{Berglund2005}%
  \BibitemOpen
  \bibfield  {author} {\bibinfo {author} {\bibfnamefont {N.}~\bibnamefont
  {Berglund}}\ and\ \bibinfo {author} {\bibfnamefont {B.}~\bibnamefont
  {Gentz}},\ }\href {\doibase 10.1209/epl/i2004-10472-2} {\bibfield  {journal}
  {\bibinfo  {journal} {Europhysics Letters ({EPL})}\ }\textbf {\bibinfo
  {volume} {70}},\ \bibinfo {pages} {1} (\bibinfo {year} {2005})}\BibitemShut
  {NoStop}%
\bibitem [{\citenamefont {L\"ofstedt}\ and\ \citenamefont
  {Coppersmith}(1994)}]{Coppersmith1994}%
  \BibitemOpen
  \bibfield  {author} {\bibinfo {author} {\bibfnamefont {R.}~\bibnamefont
  {L\"ofstedt}}\ and\ \bibinfo {author} {\bibfnamefont {S.~N.}\ \bibnamefont
  {Coppersmith}},\ }\href {\doibase 10.1103/PhysRevE.49.4821} {\bibfield
  {journal} {\bibinfo  {journal} {Phys. Rev. E}\ }\textbf {\bibinfo {volume}
  {49}},\ \bibinfo {pages} {4821} (\bibinfo {year} {1994})}\BibitemShut
  {NoStop}%
\end{thebibliography}
\end{document}